\documentclass[%
preprint,
 amsmath,amssymb,
 aps,
]{revtex4-2}
\usepackage{easyReview}
\usepackage{graphicx}
\usepackage{dcolumn}
\usepackage{bm}
\usepackage{chemmacros}
\begin{document}


\title{Overbias Light Emission From Memristive Nanojunctions\\}

\author{S. Hamdad}
\author{K. Malchow}
\author{E. Dujardin}
\author{A. Bouhelier}%
 \email{alexandre.bouhelier@u-bourgogne.fr}
\affiliation{%
 Laboratoire Interdisciplinaire Carnot de Bourgogne, CNRS UMR 6303 Université de Bourgogne Franche-Comté, 21000 Dijon, France.
}%

\author{B. Cheng}
\affiliation{%
Microelectronics Thrust, Function Hub, Hong Kong University of Science \& Technology,
Guangzhou, China
}%

\author{T. Zellweger}%
\author{J. Leuthold}%
\affiliation{%
 ETH Zurich, Institute of Electromagnetic Fields, Zurich, 8092, Switzerland
}%

\date{\today}

\begin{abstract}
A nanoscale dielectric gap clamped between two metal electrodes may undergo a large resistance change from insulating to highly conducting upon applying an electrical stress.  This sudden resistive switching effect is largely exploited in memristors for emulating synapses in neuromorphic neural networks. Here, we show that resistive switching can be accompanied by a release of electromagnetic radiation spanning the visible spectral region. Importantly, we find that the spectrum is characterized by photon energies exceeding the maximum kinetic energy of electrons provided by the switching voltage. This so-called overbias emission can be described self-consistently by a thermal radiation model featuring an out-of-equilibrium electron distribution generated in the device with an effective temperature exceeding 2000~K. The emitted spectrum is understood in terms of hot electrons radiatively decaying to resonant optical modes occurring in a nanoscale \ch{SiO2} matrix located between two \ch{Ag} electrodes. The correlation between resistive switching and the onset of overbias emission in atomic-scale photonic memristor brings new venues to generate light on chip and their exploitation in optical interconnects. Photons emitted during memristive switching can also be monitored to follow the neural activation pathways in memristor-based networks. 
\end{abstract}
\maketitle

\section{\label{sec:introduction}Introduction}

Despite its apparent simple geometry, a nanometer-wide gap separating two metal tips constitutes a unique interdisciplinary fertile ground for exploring electrical, optical, chemical and biological interactions down to the molecular or even atomic scale. On one hand, extremely large static electric fields ($> 10^8$ Vm$^{-1}$), intense electromagnetic enhancement reaching several orders of magnitude~\cite{Xu1999} and record-high nm$^3$ optical mode volumes~\cite{Buamberg_19} are now achievable with gap structures fabricated by modern patterning tools. These figures have enabled a variety of electronic and optical spectroscopies at the single molecule level~\cite{Solomon_10,Ward2008} and have fostered the emergence of new research territories such as optics in picocavities~\cite{Baumberg_22}, molecular optomechanics~\cite{Galland_21}, quantum plasmonics~\cite{aizpuruaNC12,Tame_13}, and trapping at the nanoscale~\cite{Pang2012,Berthelot_14} to name a few. 
On the other hand, electronic components integrating nanometer scale gap in their design were also crucial to the advent of novel form of computing. Memristors for instance are programmable voltage-dependant resistive devices deployed nowadays in cognitive hardware systems such as artificial neural networks, neuromorphic and reservoir computing~\cite{Li_2018,Ye_22}. In these devices, memory and processing are co-located on the same component enabling thus lower energy consumption, higher speeds and scalability~\cite{Ambrogio_19}. Memristive operation relies on resistance switching triggered by the electroformation and disruption of conductive pathways within a nanometer-scale dielectric gap~\cite{Tsuruoka_2010}. Depending on the metal and the nature of the insulating layer, charge transport occurs by an electro-chemical reduction of metal ions aggregating to conductive filaments~\cite{Tsuruoka_2010}, or by migration of mobile defects, such as oxygen vacancies~\cite{Menzel21} and nanoclusters~\cite{Tour12,Yang2014}. It is clear that such voltage-triggered modifications of the dielectric gap not only affect charge transport in the device, but also opens new control strategies over optical functionalities as well~\cite{Chen14,Brongersma16,Emboras16,Leuthold17,Xu21}. 

A particular interesting functionality enabled by ultrasmall gaps is the  generation of light by inelastic electron tunneling (IET). Such electrically-driven junction acts as a tunable optical antenna converting electrical energy into light that can be deployed for the next generation of on-chip integrated photon sources~\cite{Hecht15,Parzefall_2019}. It is not clear however if such an effect can be observed when the tunneling dielectric changes its conduction properties. In switchable resistive gap, the dielectric layer evolves dynamically and the situation drastically differs from passive tunneling barrier generally employed with electrically-driven optical antennas. The correlation between electrical and emission properties is quite challenging and the time evolution of the memristive barrier becomes an important characteristic. This is particularly the case for Ag/\ch{SiO2}/Ag configuration because the conductance of the device undergoes different transport regimes with the growth of a filament in the dielectric. For instance, defect-induced electroluminescence has been observed during the electro-formation of a planar Ag/\ch{SiO2}/Pt memristor~\cite{Cheng2022}. A complete electro-chemical formation of the filament changes the nature of transport when the conductance of the device $G$ increases close to the quantum point contact $G_0=2e^2/h=7.75\cdot 10^{-5}$ S, where $h$ is the Planck constant and $e$ is the elementary charge. Near this conductance value even a small atomic reorganisation of the gap induces a significant variation of the conductance. Indeed, electrons injected and transported through the gap may take different pathways (tunneling, hopping, ballistic, etc.~\cite{Menzel21}) which are all determined by the dielectric properties of the gap and the bias value. This aspect sets the importance of a precise and accurate control of the applied electrical constraint as it defines the irreversible historicity of the memrisitive junction. 

In this paper we identify yet another light emission mechanism occurring in a memristive Ag/\ch{SiO2}/Ag junction that is departing from inelastic electron tunneling or from radiative recombinations of charge carriers. Specifically, we show that when the device undergoes resistive switching to a conductance value slightly below $G_0$ light is released in the so-called overbias emission regime.  In this mode of operation, the energy $h\nu$ of the photons emitted exceeds the kinetic energy of a single electron $eV_{\rm b}$ traversing the device. $\nu$ is the frequency of light and $V_{\rm b}$ is the voltage drop at the two terminals of the device. Overbias emission has been repeatedly observed in various types of metal contacts ranging from discontinuous islands~\cite{Tomchuk66}, electromigrated nanowires~\cite{Buret2015,Cui20}, mechanical break junctions~\cite{Dumas16} and scanning tunneling experiments~\cite{welland02,Berndt17}, but has not been associated to the memristive change of a dielectric matrix. A memristor has several key advantages compared to other geometries. The device holds a dual functionality whereby electrons and photons are controlled by the same atomic-scale system, its response can be cycled and it offers potential for integration and scalability.

\section{\label{sec:fabrication}Fabrication and Electroformation of the device}
\begin{figure}[h]
    \centering
    \includegraphics[width=1\linewidth]{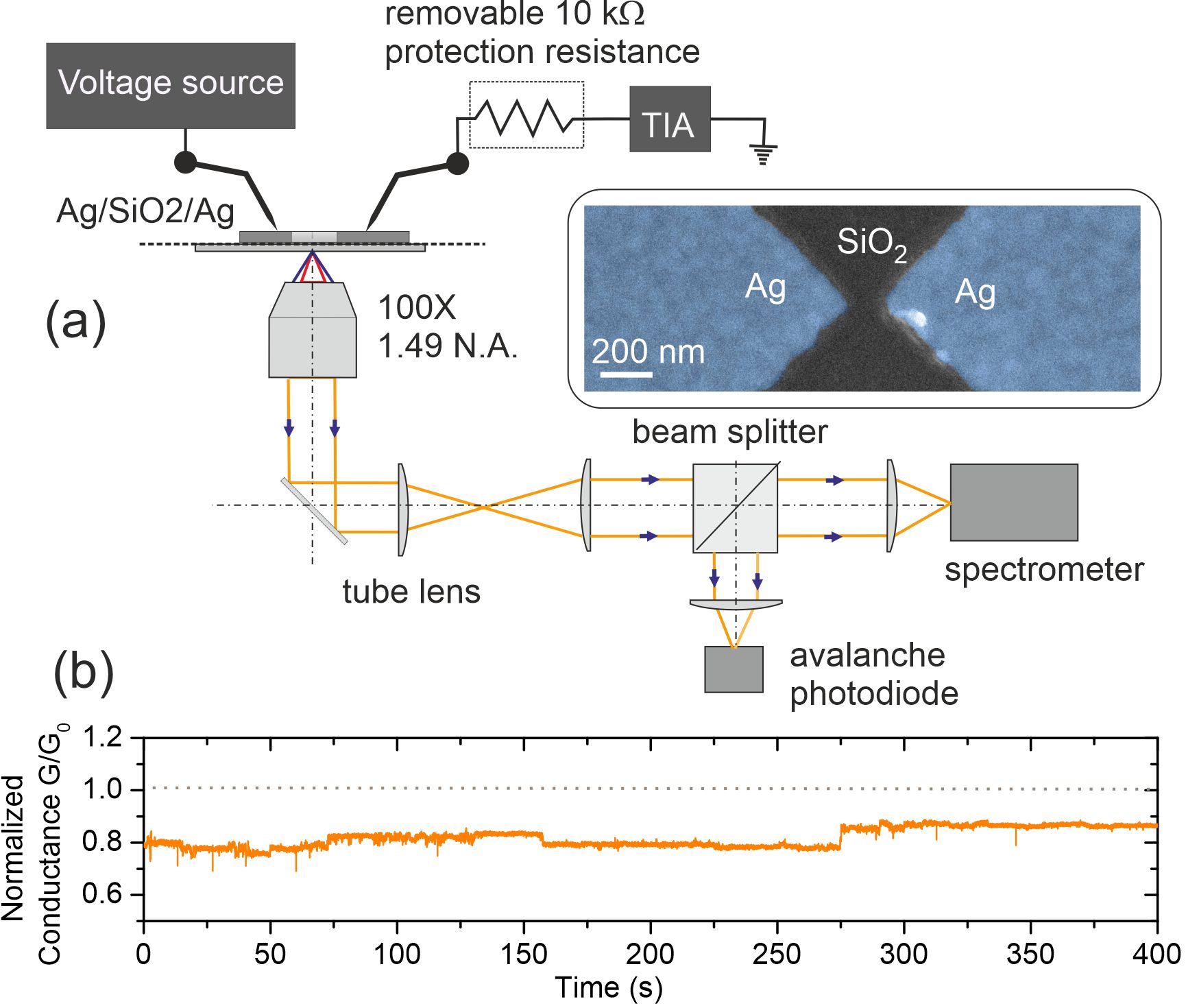}
    \caption{ (a) Schematic representation of the experiments including the electrical excitation and signal recovery with transimpedance amplifier (TIA) as well as the optical detection. Inset: SEM image of a typical in-plane Ag/\ch{SiO2}/Ag junction. Ag is colorized with a blue hue and a sacrificial conducting layer has been deposited to allow for electron imaging. (b) Time trace of the device's normalized conductance $G/G_0$ during the final stage of the activation process.}
    \label{fig:figure1}
\end{figure}

The device geometry discussed here is constituted of two in-plane 70 nm-thick silver electrodes thermally deposited on an ultrathin Cr adhesion layer ($\sim 1$~nm) on top of a \ch{SiO2} glass coverslip. The electrodes are terminated by a tapered section forming a $90^\circ$ angle and are separated by a gap of about 60~nm. The structure is realized by electron beam lithography complemented by metal deposition and a liftoff process. The inset of Fig.~\ref{fig:figure1}(a) shows a scanning electron micrograph (SEM) of a nominal device. 

The device is operated under ambient conditions and its electro-optical responses are measured using the setup schematically presented in Fig.~\ref{fig:figure1}(a). The junction is electrically driven by a voltage source applied on one electrode while the other is connected to a transimpedance amplifier (TIA). The TIA accurately measures the current flowing in the device. During the initial characterization steps described below, we introduce a protection resistor (10 k$\Omega$) to limit the maximum current. Light emission produced by the device is collected by a high numerical aperture objective and is directed to an avalanche photodiode for recording the photon count rate and to an imaging spectrometer for analyzing the spectral content.

For a 60~nm gap size and biases of a few volts, the dielectric spacer is too large to enable electrons to tunnel from one electrode to the other; the device is in a high resistive state (HRS). To initiate memristive switching to a low resistive state (LRS), we repeatedly stress the device by applying voltage ramps of increasing amplitudes.
During this electroformation phase, a conduction path dominantly composed of a mixture of silver aggregates builds up between the electrodes~\cite{Wang17,Patel19} and a low current ($\sim$~nA) flowing can be registered.  Once the junction is electroformed, reversible resistive switching from HRS to LRS occurs when the first filamentary conductive path connects these discontinuous nanoclusters. This step is critical as it constitutes the historicity of the junction, the morphology of the conduction path and the electrical characteristics at which the device will operate. Figure~\ref{fig:figure1}(b) is a time trace of the normalized conductance of the device $G/G_0$ recorded during the last electroformation stress. The conductance, estimated by taking into account a constant driving voltage (400 mV) and the voltage drop at the protection resistance, is here stabilized slightly below $G_0$. This suggests that current transport is taking place through an incomplete or imperfect filament formation. The step-like evolution of $G/G_0$ observed at different moments (e.g., $t=72$~s, 155~s or 275~s) indicates a dynamic reorganisation of the current pathways either by a migration of species or ionization of charge traps. This LRS is unstable and relaxes to a HRS when the electrical stress is stopped. This is an asset in our study because we intend to capture the emission property of a volatile memristive device repeatedly switching at high conductance value ($\sim G_0$). 

\section{\label{sec:electroOptical}Electro-optical properties of the device}

The dynamic of filament formation is typically triggered by two crucial parameters which are the applied voltage and the growth time. To avoid the formation a complete silver bridge, which would be materialized by a diffusive transport regime ($G \gg G_0$) and no photon emission, we operate the electroformed device with sequences of 1.24~V voltage pulses. Each sequence is composed of 100 pulses with 100 ms duration and a period of 500 ms. The rising edge of the pulse 1.3~$\mu$s. The volatile nature of the memristor causes the system to relax (LRS $\Rightarrow$ HRS) during the time the voltage is off, thus preventing a metallic contact to form between the two electrodes. Joule heating during transport may also contribute to destabilize the filament and prevent the complete growth because the protection resistance has been removed after the electroformation phase of the device in order to have all the applied voltage dropped at the gap.

\begin{figure}[h]
    \centering
    \includegraphics[width=1\linewidth]{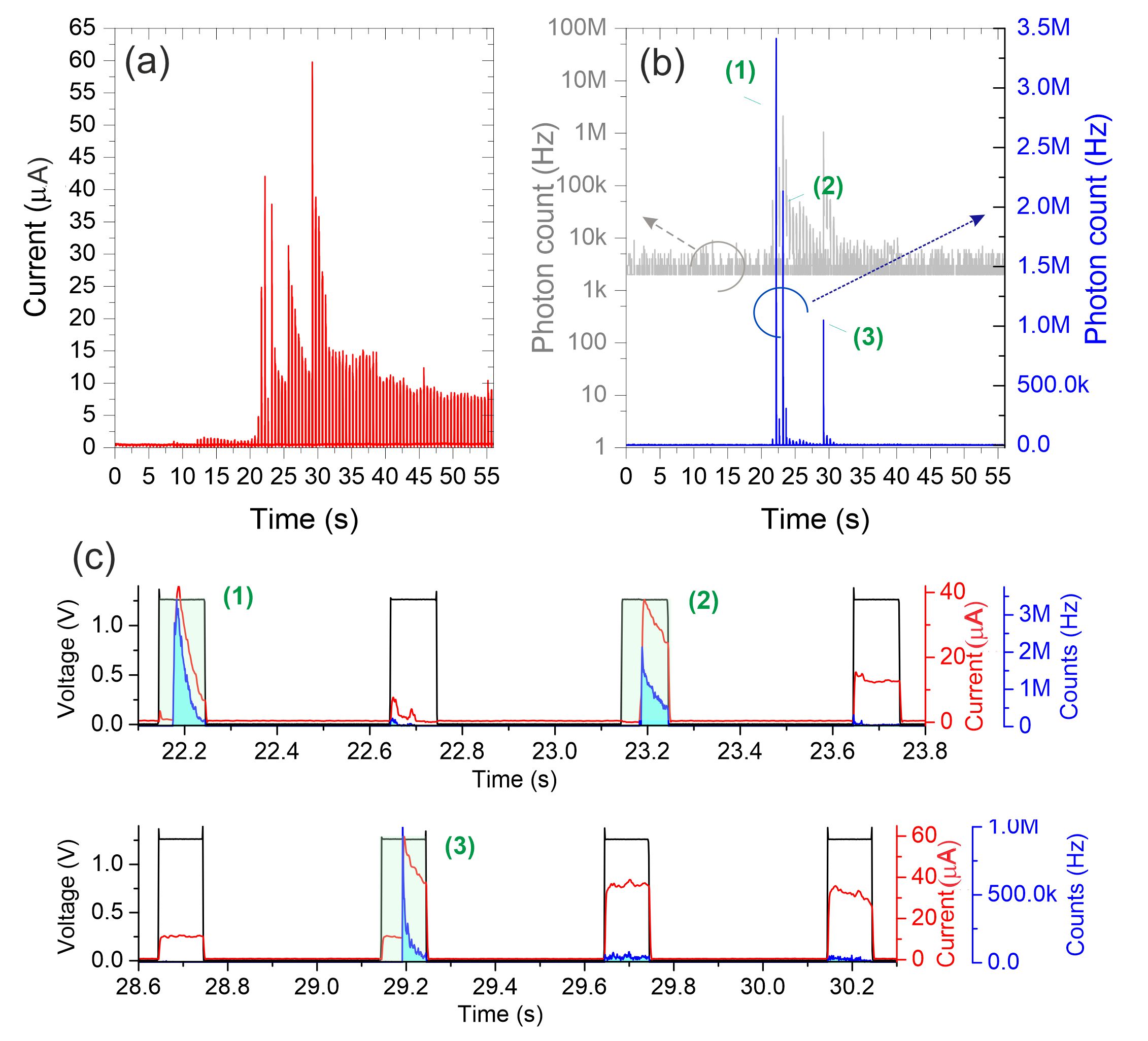}
    \caption{(a) and (b): time trajectories of the current and photon counts acquired during the application of a train of voltage pulses ($V_{\rm b}=1.24$~V, not shown for the sake of clarity). Photon counts are represented on a linear (blue) and logarithmic (grey) scales. (c) temporal sequence featuring the three most intense events (shaded pulses) labeled (1), (2) and (3) in (b).}
    \label{fig:figure2}
\end{figure}

Figure~\ref{fig:figure2}(a) shows a typical behavior of the current acquired during a $\sim$ 1 min long pulse sequence. During the first 12~s, the device remains in a HRS and the current is t the noise floor, here about 400 nA. From $12~{\rm s}<t<20.7$~s, a small current ($\sim 1.5~\mu$A) indicates a progressive build up of the conduction paths. Resistive switching depends of the number of available Ag ions present in the dielectric layer. At the beginning of the sequence, too few ions are present to switch the device. At $t>21$~s, the device undergoes a clear resistive switching to a LRS at each voltage pulse. We observe a transient response with the occurrence of fluctuating current peaks up to $t\sim32$~s. The LRS current then stabilizes to about 10 $\mu$A during the rest of the sequence to a stationary LRS of $G\approx0.1G_0$. During the transient phase, the APD records a strong optical activity that eventually vanishes when the memristor switches to a stable LRS. This is clearly depicted in Fig.~\ref{fig:figure2}(b) showing the time trajectory of the photon counts. To appreciate even the dimmest light emission, we also plot the count rate in logarithmic units (grey curve, right axis). The photon count is particularly large for three switching pulses at $t=22.18$~s, $t=23.18$~s, and $t=29.29$~s, labeled (1), (2), and (3) respectively. There, the count rates are in excess of 1~MHz, a value nearly two orders of magnitude higher than typical single molecule rates (kHz) detected within the same diffraction-limited volume~\cite{michaelis2000}. Figure~\ref{fig:figure2}(c) shows a time extract of the transient phase where most of the photons are detected. The intense events are highlighted by shaded areas and are occurring at peaking conductance values of  $G \sim 0.4 G_0$, $G\sim 0.35 G_0$ and $G\sim 0.6 G_0$, respectively. Note that both current and photon counts have a delayed response of $\sim$ 30~ms with respect to the rising edge of the voltage pulse. The delay is understood from the intrinsic dynamics of the conductive path formation when the bias is applied and pulse-to-pulse variation of the density of ions available.

There is an obvious timing correlation between luminous events and current peaks linked to the unsteady LRS. This is seen within the shaded pulses of the time trace in Fig.~\ref{fig:figure2}(c) where a rapid drop of the current is observed after the emission is triggered.Photons are clearly produced during a dynamic process taking place within the device. A detailed explanation of the current fluctuations is still unclear. A series of mechanisms may contribute to the effect including atomic reshaping of the filament and suppression of conduction channels by charged traps. On the contrary, a stable and somewhat lower LRS (\emph{i.e.} a weakly fluctuating current during the pulse) triggers little or no optical activity from the device.  In these pulses, the onset of the electron flow is concurrent to the rising edge of the voltage pulse (no delayed response). This is seen for instance in Fig.~\ref{fig:figure2}(c) for pulses at $t=23.65$~s, $t=28.65$~s, or $t=29.65$~s. Probably because of a persistent density of ions~\cite{Cheng_19}, resistive switching is therefore occurring at a lower voltage than the shaded pulses. The times traces unambiguously show that these resistive switching cycles are dark.

\begin{figure}[h]
    \centering
    \includegraphics[width=1\linewidth]{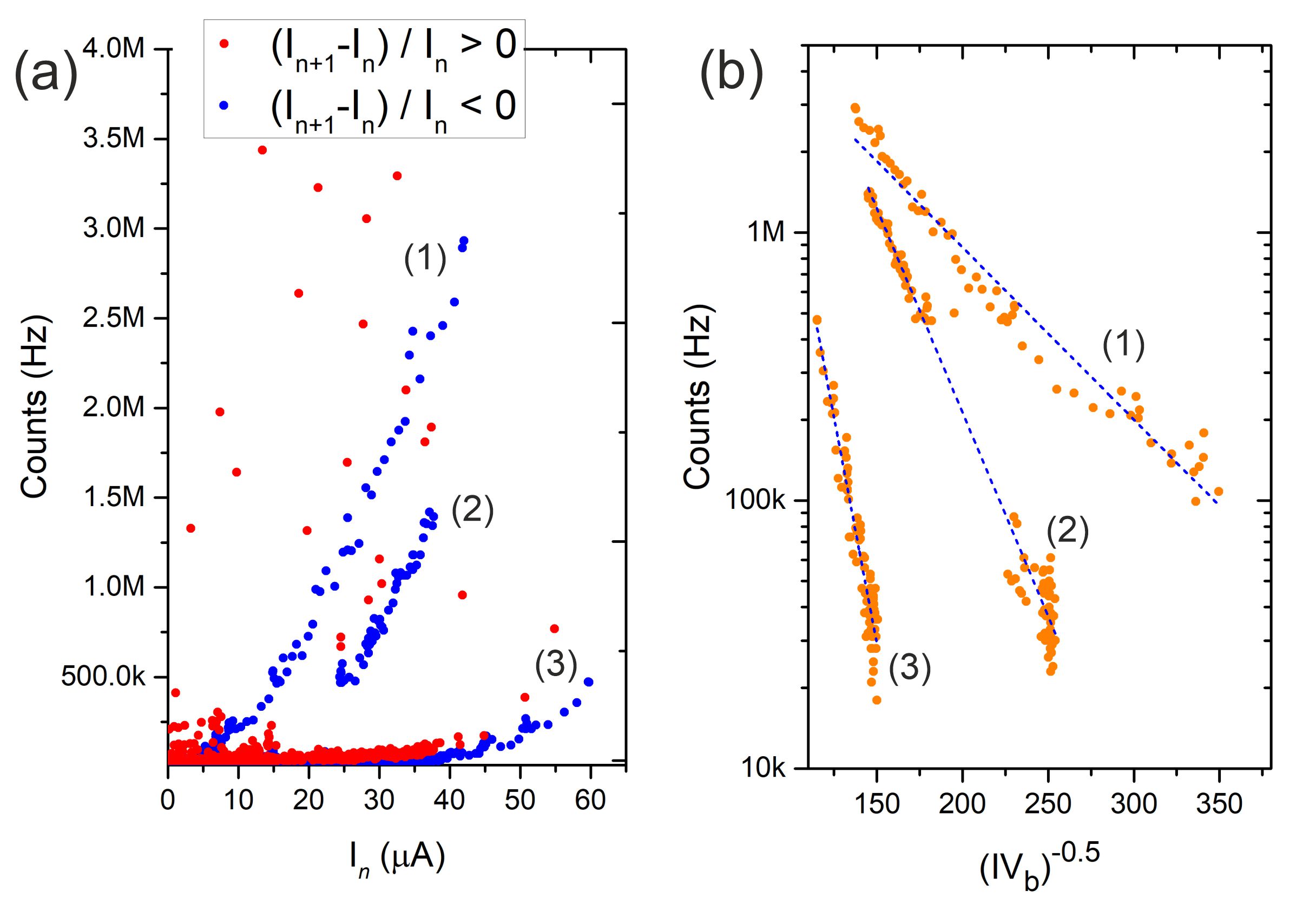}
    \caption{(a) Separate contributions of the relative increase (red squares) and the relative decrease of the current (blue points) to the photon count rate. The three branches forming the blue data set are linked to the events (1) to (3) of Fig.~\ref{fig:figure2}. (b) Logarithmic plot of the count rate as a function of the inverse square root of the dissipated power for each of the three branches observed in (a).}
    \label{fig:figure3}
\end{figure}

To investigate the correlation between current fluctuations and photon generation, we examine their dependence by analyzing the variation $\Delta I = I_{(n+1)}-I_n$ inferred around each $I_n$ values, where $n$ is the current bin number in the time trace.  Results are shown in Fig~\ref{fig:figure3}(a). The plot separates the contribution to the light emission of a relative increase of the current ($\Delta I/I>0$) to that of a relative decrease of the current ($\Delta I/I<0$). 

For variation leading to an increase of the current, the device generally has a very low optical activity (red $\blacksquare$). There are a few scattered points associated with the detection of intense signal but there is not clear trend with the current. We can conclude that the photon count is not related to the number of electrons flowing in the device for $\Delta I/I> 0$. On the contrary, for negative fluctuations of the current, three branches clearly appear in the graph (blue $\bullet$). These branches are related to the sequence of the three most luminous events observed on the time trace of Fig.~\ref{fig:figure2}. A series of observations can be made by analyzing Fig.~\ref{fig:figure3}(a). First, the correlation plot reveals the nonlinear nature of the photon signal since the emission is only detected after a current threshold. The onset of current triggering photon emission increases with each branch with: $I\sim3~\mu$A for (1), $I\sim10~\mu$A for (2) and $I\sim35~\mu$A for (3) suggesting an evolving response. Second, Fig.~\ref{fig:figure3}(a) also shows that the absolute amplitude of the current is not the proper parameter that sets the rate at which the devices emits: a similar current gives different photon counts for the three branches. These two observations taken together suggest an underlying sensibility of the emission to the peculiarity of the current channel formation at the atomic level as well as the history of the device. Third, the correlation between photon counts and negative $\Delta I/I$ clearly indicates that the process of light emission affects the electron transport within the device. We shall come back to this point later in the discussion. 

These series of measurement and dependencies were observed on multiple occasions. We show in the appendix a set of data acquired during a second run obtained with the same device which is complementing and confirming the trends discussed in Fig.\ref{fig:figure2} and Fig.\ref{fig:figure3}. 

\section{\label{sec:overbiasl} Overbias emission}
Let us discuss now the origin of the light signal. Our starting point is the following: the avalanche photodiode used in the experiment has a detection efficiency peaking at 1.77 eV (700 nm) which rapidly vanishes to $<2$\% below 1.2 eV.  Any emission mechanism promoted by a single electron process would therefore be emitted in the nearly blind spectral region of the detector. The optical activity registered in the time traces of Fig.~\ref{fig:figure2} suggests that the memristive device releases photons in an overbias emission regime. Opposite to single electron process producing photon energies smaller than the voltage applied ($ h\nu \leq eV_{\rm b}$), overbias light emission (or sometimes referred as above-threshold or anomalous) requires the cumulative contributions of multiple electrons. 

Different mechanisms have been proposed to explain overbias emission including thermal broadening of the Fermi-Dirac distribution~\cite{pechou98}, coherent multi-electrons scattering~\cite{schull09,Schneider13,belzig14,Nitzan15,Venkataraman20} and decay of hot carriers~\cite{Buret2015,Dumas16,Zhu2020}. For interacting systems with characteristic dimensions smaller than the electron mean-free path (about 30 nm to 60 nm for noble metals at room temperature~\cite{Chopra63,Sze64}) and high pumping conditions, the effective temperature of the electron subsystem $T_{\rm e}$ may be set by the electrical power dissipated. This is the main line of argumentation put forward by Fedorovich, Naumovets and Tomchuk in their pioneer work on electrically connected discontinuous island film~\cite{Federovich00}, where the power dissipated in the system results to steady-state non equilibrium electron and lattice temperatures ($T_{\rm e} \neq T_{\rm L}$). Under this circumstance, the electron temperature is written $T_{\rm e} =\sqrt{\alpha IV_{\rm b}}/k_B$ and can reach values greatly exceeding the lattice temperature which remains virtually unchanged $(T_{\rm e} >> T_{\rm L})$. Here $k_B$ is Boltzmann constant and $\alpha$ is an empirical factor which depends on the exchange interaction between electrons and lattice phonons~\cite{Buret2015,Zhu2020}. 

In this picture, the net energy gain observed for photon in the regime of overbias ($ h\nu \geq eV_{\rm b}$) is attributed to inelastic scattering of the hot electrons with the physical boundaries giving rise to a Bremsstrahlung emission~\cite{Buret20}. However, thermal Bremsstrahlung of colliding hot electrons cannot be distinguished from a spontaneous decay of out-of-equilibrium electrons because both mechanisms obey the same energy dependence in the detected energy range~\cite{Ghisellini,Buret20}. The problem can be reduced to a simple blackbody thermal radiation model~\cite{Federovich00} with spectral density $U(V_{\rm b},\nu)$. 

\begin{equation}
\label{eq:BB}
  U(V_{\rm b},\nu)\propto \rho(\nu)h\nu\ e^{- 
  h\nu/\sqrt{\alpha IV_{\rm b}}}
\end{equation}

where $\rho(\nu)$ is the local density of optical states. Equation~\ref{eq:BB} can be reformulated to
\begin{equation}
\label{eq:lnBB}
  \ln[U(V_{\rm b},\nu)]\propto \ln[\rho(\nu)h\nu]-\frac{h\nu}{\sqrt{\alpha IV_{\rm b}}}
\end{equation}

Equation~\ref{eq:lnBB} highlights the expected nonlinear dependence of the light signal with the electrical power $IV_{\rm b}$ dissipated in the device. Figure~\ref{fig:figure3}(b) shows the photon count rate in logarithmic scale as a function of the parameter $1/\sqrt{IV_{\rm b}}$ extracted from the three branches of Fig.~\ref{fig:figure3}(a). Clearly, the data associated to each $i$-branch follow the expected linear dependence of the heated electron gas model (Eq.~\ref{eq:lnBB}) with slopes given by $\beta_{(i)}=-h\nu/\sqrt\alpha_{(i)}$. The dashed blue curves are linear fits to the data providing  $|\beta_{(1)}|=15\times 10^{-3}\ {\rm W}^{1/2}$, $|\beta_{(2)}|=35\times 10^{-3}\ {\rm W}^{1/2}$ and $|\beta_{(3)}|=78\times 10^{-3}\ {\rm W}^{1/2}$. These slopes only give an estimate because the photon count integrates the light emitted over the whole detection energy. The different $\beta_{(i)}$ values are related to the particularity of the atomic geometry which dictates the thermalization process of electrons. 

We confirm that our results are in line with this out-of-equilibrium regime by two conclusive observations. Assuming a photon energy at the peak efficiency of the detector, we can easily estimate an effective electronic temperature using the relation $T_{\rm e} =\sqrt{\alpha IV_{\rm b}}/k_B$. If we take for instance the second branch $i=2$, we find that $T_{\rm e}$ is about 4000~K at the highest count rate and 2300~K at the lowest. This range of $T_{\rm e}$ is consistent with electronic temperatures reported in electromigrated junctions and quantum point contacts operated under comparable electrical excitation and emitting also in overbias regime~\cite{Buret2015,Zhu2020,Buret20}.  In our planar memristive junction, there is a clear interdependence between the current transported in the device and the light emission. During the pulses providing the brightest events displayed in the time trace of Fig.~\ref{fig:figure2}(c), current and photon counts share a similar decaying behavior. This correlative feature suggests that the thermalization process of hot carriers affects the transport, probably by gradually disrupting the filamentary pathway. In turns, the electrical power dissipated in the system reduces, and $T_{\rm e} $ decreases concomitantly to its the overbias electromagnetic signature. 

\begin{figure} [h]
    \centering
    \includegraphics[width=1\linewidth]{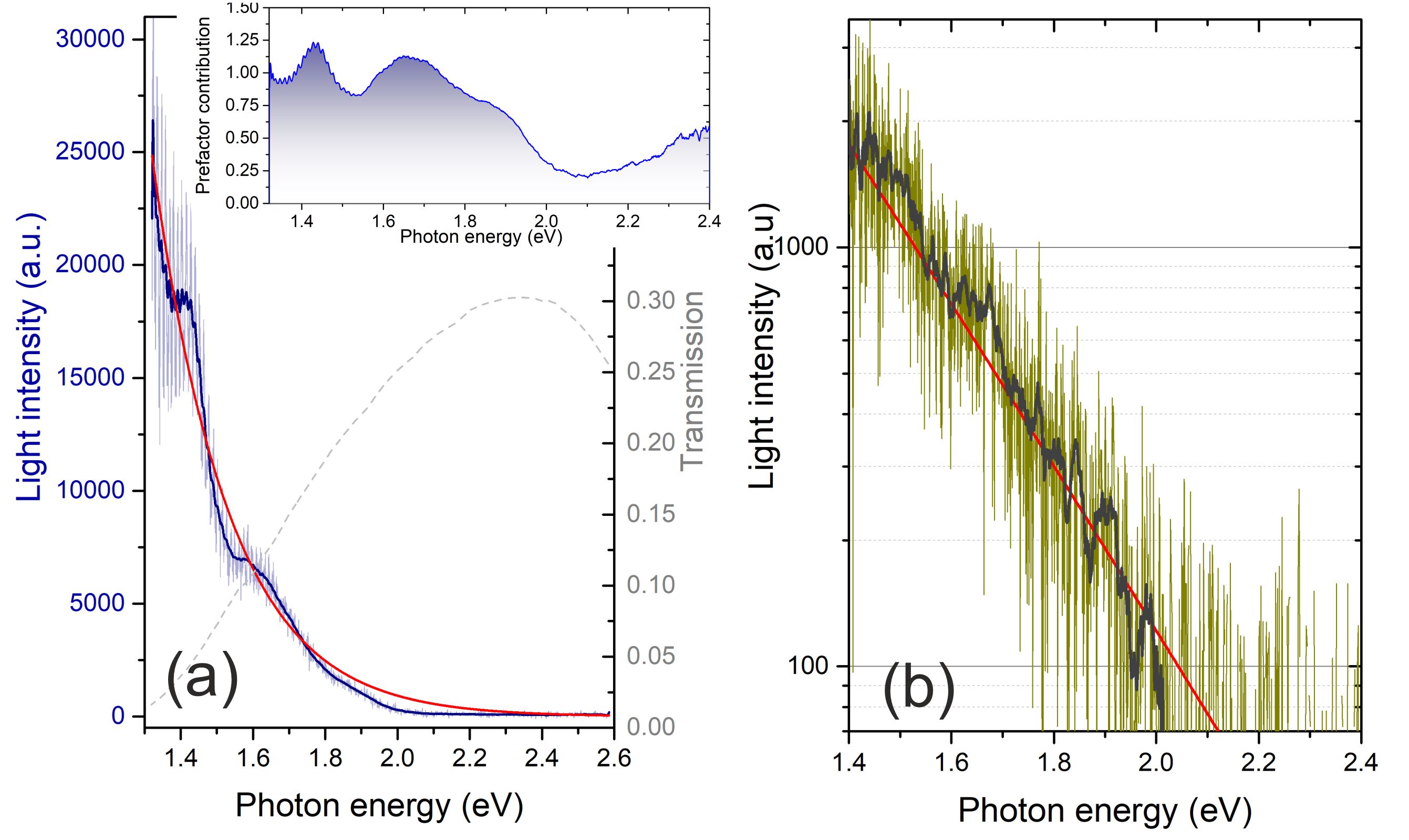}
    \caption{(a) Overbias emission spectrum of the second luminescent event (2). The light blue line (left axis) is the corrected spectrum taking into account the spectral efficiency of the detection (grey curve, right axis). The dark blue line is an adjacent averaging. The red line is a fit of the data using Eq.~\ref{eq:BB} with a dispersionless prefactor $\rho$. Inset: spectral contribution of the prefactor obtained after normalizing the experimental spectrum by Planck's contribution. (b) Overbias emission spectrum of the third luminescent event (3) registered in Fig~\ref{fig:figure2}(b).  The red line is a fit of the data using Eq.~\ref{eq:BB} with a dispersionless prefactor $\rho$ and $T_e=2300$~K. Semi-logarithmic scale. }
    \label{fig:Figure4}
\end{figure}

The second experimental evidence confirming the origin of the light released upon resistive switching is its spectral content. Figure~\ref{fig:Figure4}(a) shows the spectrum acquired during the second luminous event recorded in the time trace of Fig.~\ref{fig:figure2}(c). The spectrum is corrected for the overall collection efficiency of our instrument, shown in the dashed grey curve, using the objective and spectrometer's specifications. The correction explains the enhanced noise at low energies where the silicon detector has a vanishing efficiency. The corrected spectrum features a characteristic decay tail consistent with an overbias emission regime. The light recorded during the switching pulse clearly violates the quantum cutoff set by single electron process ($h\nu = eV_{\rm b}$) because the maximum photon energy ($\sim$2 eV) largely exceeds the kinetic energy of carriers (1.24 eV). The spectrum cannot be explained by any linear electroluminescent colored centers which may be present within the dielectric matrix and emitting over a defined energy range~\cite{Cheng2022} and is also very different from coherent multielectron emission mechanism reported with STM where clear spectral signatures at $neV_{\rm b}$ associated with $n$-electron processes~\cite{schull09,Natelson22}.  

The spectrum shows a series of shoulders, which under the light of Eq.~\ref{eq:BB}, can be reasonably interpreted by the spectral evolution of the local density of optical states $\rho(\nu)$. Photon emission is here triggered by hot electrons colliding with the boundaries of a complex optical environment sustaining a rich surface plasmon landscape. Energy dispersion of $\rho(\nu)$ is thus expected in such composite nanoscale structures~\cite{Buret2015,Cui20}. An independent confirmation of these resonances is a daunting task because the characteristic length at which the interaction occurs is beyond the capabilities of standard optical spectroscopies. Nonetheless, we tentatively extract from the spectrum an effective temperature of the electron bath giving rise to the overbias light emission. We fit the averaged data points (blue line in \ref{fig:Figure4}(a)) with Eq.~\ref{eq:BB} considering a constant prefactor $\rho$ and replacing $\sqrt{\alpha IV_{\rm b}}$ by the standard temperature dependence $k_BT_{\rm e}$, with $T_e$ used as a free parameter. The fit is shown as the red line with $T_e=2127$~K. Conservation of energy requires that $k_BT_e \leq eV$. Using the electron temperature extracted from the fit of Fig.~\ref{fig:Figure4}(a), we find that $k_BT_{e}/eV$ is about $0.14$. The value is consistent with a previous report of overbias emission obtained with STM point contacts~\cite{welland02}. The inferred electron temperature is a little bit lower than the range of effective $T_e$ inferred by analyzing Fig.~\ref{fig:figure3} where $2300~\rm{K} <T_e<4000$~K. The discrepancy is probably linked to the accuracy of the estimation deduced from a spectrally integrated measurement and also to the imprecision of the efficiency curve deduced from the manufacturer datasheets and used to correct the raw spectrum. 

Having inferred the effective temperature of the electron sub-system, we extract the contribution of the prefactor $\rho(\nu)$ by normalizing the experimental spectrum by $h\nu\ e^{- h\nu/k_BT_e}$. The spectral landscape of the prefactor is shown in the inset of Fig.~\ref{fig:Figure4}(a). Weak resonances can be seen when $\rho(\nu)>1$ at 1.42 eV and 1.66 eV which are reminiscent of the underlying geometry. Because the device is reconfiguring itself at every voltage pulse (volatile switching), the spectral position and strength of the resonances are likely to vary from pulse to pulse. This is illustrated by the spectrum of the third event displayed in Fig.~\ref{fig:Figure4}(b). This event was the weakest of the three, but the spectrum clearly confirms the overbias nature of the emitted photons. The spectrum is displayed in semi-logarithmic scale to visualize the exponential behavior expected from Eq.~\ref{eq:lnBB}. The fit gives an effective electronic temperature at 2300 K, a slightly higher value compared to the second event despite a lower emitted photon rate. This may be understood from the shape of the spectrum where no resonances are present to increase the radiation efficiency of the thermal emission.
Unfortunately, the first event was not captured by the spectrometer because of a synchronisation problem with the voltage pulses.

As a point of discussion, we note that in a 2020 paper, E. Fung and L. Venkataraman questioned this hot electron picture on the basis of data obtained by scanning tunneling microscopy experiment. They proposed an alternative emission mechanism involving multi-electron interactions~\cite{Venkataraman20} and were able to reproduce the 
$\sqrt{\alpha IV_{\rm b}}$ dependence by considering a third order process. It is not clear if this mechanism can be applied to the present case since the range of electrical power explored is significantly greater. We note that crossover between coherent multielectron photon emission and hot-carrier radiation has been recently observed for in-plane devices placed in a cryostat where hot electron radiation where shown to be prevalent with noble metal plasmonic systems~\cite{Natelson22}.  In our Ag memristive junction operated at ambient conditions, the density of optical states is dominated by the surface plasmon spectral landscape, favoring thus a hot electron overbias emission regime. Additionally, because of the very small interaction area, energy dissipation is mainly guided by electrons interacting with the surface rather than electron-phonon interactions~\cite{Buret20} and we were not able to observe co-existing emission regimes. 

\section{\label{sec:conclusion} Conclusion}
We report that resistive switching in filament-type memristive Ag/\ch{SiO2}/Ag junction may be associated with an overbias light emission. In this particular emission regime, the photons released by the device are emitted by the radiative decay of out-of-equilibrium electrons brought at elevated temperature ($>$ 2000~K). This is made possible by operating the memristor with large current ($>10~\mu$A) and relatively low pulse voltages (100 ms, $\sim$ 1V) near $G_0$ conductance. The overbias spectrum is well described by a blackbody thermal radiation model that combines the Planck term with a local density of states reflecting the underlying surface plasmon landscape as well as material properties. These dependencies suggest that structuring the local electromagnetic environment of the memristor near the resistive switching region may help at engineering the emission spectrum and increase the efficiency of the process. 

Overbias emission triggered by the passage from high resistive state to a low resistive state does not occur systematically for every switching voltage pulse. This underpins the drastic influence of the diffusion dynamics of the Ag atoms reconfiguring the current pathway within the dielectric medium. We also stress that the stability of the low resistive state is strongly reduced when hot carriers are generated. Controlling and stabilizing the volatile condition triggering overbias photons will bring an optical functionality to the next generation of photonic memristors. Together with their ability at mimicking neurons, such dual device can also act as integrated atomic sources of broadband radiation with temporal bandwith only limited by the electron-phonon energy exchange time.

\begin{acknowledgments}
This work has been partially funded by the French Agence Nationale de la Recherche (ANR-20-CE24-0001 DALHAI and ISITE-BFC ANR-15-IDEX-0003), the EIPHI Graduate School (ANR-17-EURE-0002), and the European Union through the PO FEDER-FSE Bourgogne 2014/2020 programs. ETH acknowledges support of the Werner Siemens-Stiftung (WSS). Device characterization was performed at the technological platforms SMARTLIGHT and ARCEN Carnot with the support of the French Agence Nationale de la Recherche under program Investment for the Future (ANR-21-ESRE-0040), the R\'egion de Bourgogne Franche-Comt\'e, the CNRS and the French Renatech+ network.
\end{acknowledgments}

\appendix*

\section{Overbias emission analysis from another run}

\begin{figure}[h]
    \centering
    \includegraphics[width=0.8\linewidth]{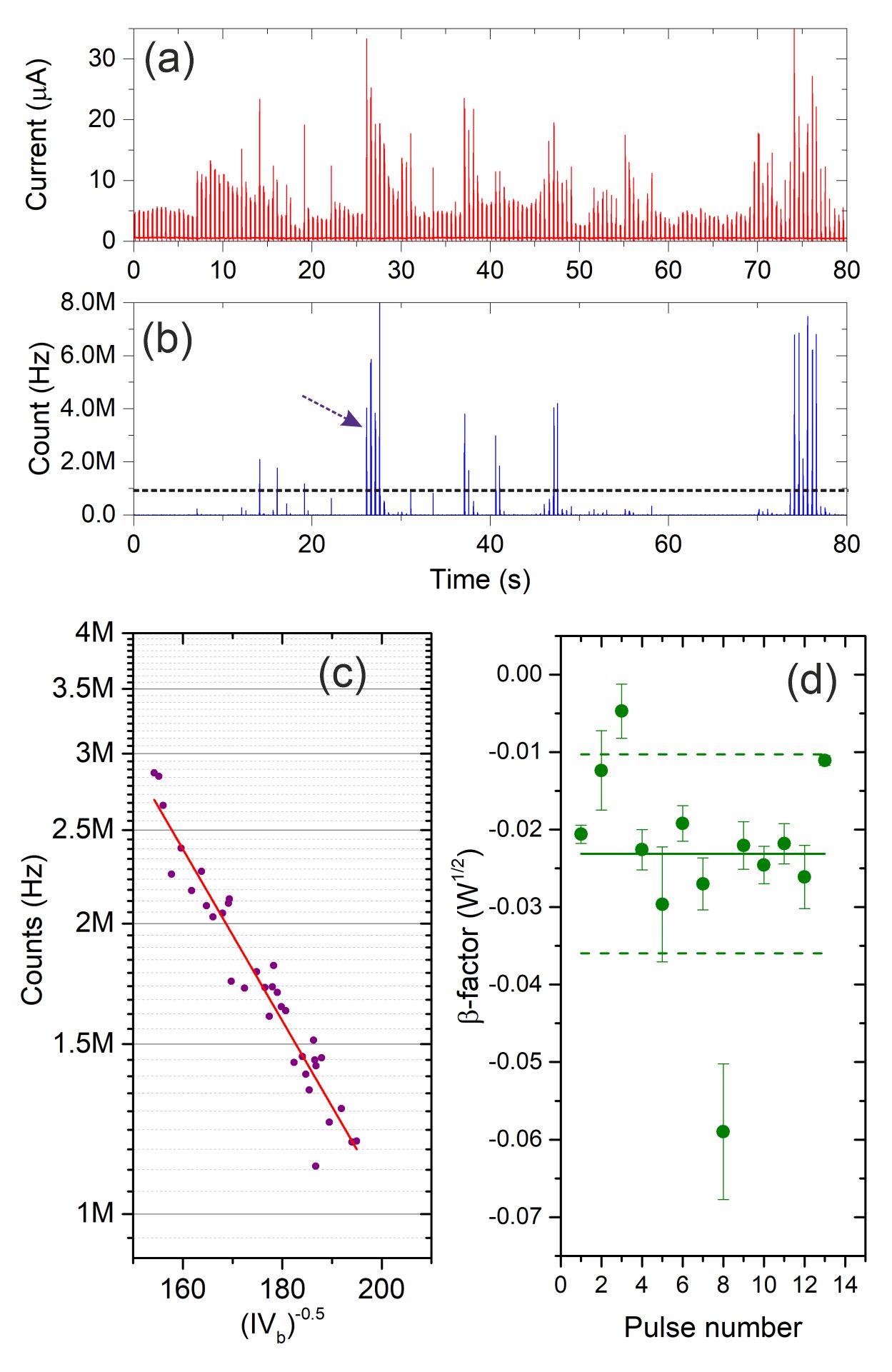}
     \caption{ (a) and (b) Time traces of the measured current and the detected count rate during a second run with the same pulse sequence (no shown).  (c) Logarithmic plots of the spectrally integrated count rate as a function of $1/\sqrt{IV_{\rm b}}$ for the event labeled by an arrow in (b). The slope of the linear fit (red line) provides $\beta$ for this event. (d) Values of $\beta$ estimated from events brighter than 1MHz (dotted horizontal line in (b)). The solid line is the mean value and the dotted lines are 1$\sigma$ deviation.}
    \label{fig:annex}
\end{figure}

To confirm that the overbias emission can be observed systematically despite the stochastic dynamic of the atomic configuration, we show in Fig.~\ref{fig:annex} the results of a second excitation sequence obtained from the same device under the same biasing conditions. The time traces of the current (Fig.~\ref{fig:annex}(a)) and the photon count rate (Fig.~\ref{fig:annex}(b)) display a strongly fluctuating behavior between all switching pulses (voltage signal not shown). Again, periods of more stable LRS such as $t<10$~s, $60~\rm{s}<t<70$~s, or $80~\rm{s}<t<105$~s do not provide any appreciable photon counts. On the contrary, current spikes associated with a re-ogarnization of the filamentary path are generally responsible for an overbias activity. Out of the many spikes present, we select an event where $G\sim 0.2~G_0$ labeled with an arrow in Fig.~\ref{fig:annex}(b). We then conduct the same analysis as in Fig.~\ref{fig:figure3} to confront the data with the thermal radiation model. The results are plotted in Fig.~\ref{fig:annex}(c) in a semi-logarithmic plot and unambiguously confirms the previous findings. The linear dependence to $\sqrt{IV_b}$ gives a slope $|\beta|=20\times 10^{-3}$~W$^{1/2}$. We then infer $\beta$ for all the events with a count rate greater than 1 MHz (dashed horizontal line in  Fig.~\ref{fig:annex}(b)). The results are plotted in Fig.~\ref{fig:annex}(d). A couple of pulses are outside the standard deviation $\sigma$, but most of the $\beta$ are near the mean value (solid line). This suggests that the device intermittently recovers the condition to create hot electronic subsystems despite a stochastic atomic re-organisation of the current channels.

\bibliography{references}

\begin{thebibliography}{48}%
\makeatletter
\providecommand \@ifxundefined [1]{%
 \@ifx{#1\undefined}
}%
\providecommand \@ifnum [1]{%
 \ifnum #1\expandafter \@firstoftwo
 \else \expandafter \@secondoftwo
 \fi
}%
\providecommand \@ifx [1]{%
 \ifx #1\expandafter \@firstoftwo
 \else \expandafter \@secondoftwo
 \fi
}%
\providecommand \natexlab [1]{#1}%
\providecommand \enquote  [1]{``#1''}%
\providecommand \bibnamefont  [1]{#1}%
\providecommand \bibfnamefont [1]{#1}%
\providecommand \citenamefont [1]{#1}%
\providecommand \href@noop [0]{\@secondoftwo}%
\providecommand \href [0]{\begingroup \@sanitize@url \@href}%
\providecommand \@href[1]{\@@startlink{#1}\@@href}%
\providecommand \@@href[1]{\endgroup#1\@@endlink}%
\providecommand \@sanitize@url [0]{\catcode `\\12\catcode `\$12\catcode
  `\&12\catcode `\#12\catcode `\^12\catcode `\_12\catcode `\%12\relax}%
\providecommand \@@startlink[1]{}%
\providecommand \@@endlink[0]{}%
\providecommand \url  [0]{\begingroup\@sanitize@url \@url }%
\providecommand \@url [1]{\endgroup\@href {#1}{\urlprefix }}%
\providecommand \urlprefix  [0]{URL }%
\providecommand \Eprint [0]{\href }%
\providecommand \doibase [0]{http://dx.doi.org/}%
\providecommand \selectlanguage [0]{\@gobble}%
\providecommand \bibinfo  [0]{\@secondoftwo}%
\providecommand \bibfield  [0]{\@secondoftwo}%
\providecommand \translation [1]{[#1]}%
\providecommand \BibitemOpen [0]{}%
\providecommand \bibitemStop [0]{}%
\providecommand \bibitemNoStop [0]{.\EOS\space}%
\providecommand \EOS [0]{\spacefactor3000\relax}%
\providecommand \BibitemShut  [1]{\csname bibitem#1\endcsname}%
\let\auto@bib@innerbib\@empty
\bibitem [{\citenamefont {Xu}\ \emph {et~al.}(1999)\citenamefont {Xu},
  \citenamefont {Bjerneld}, \citenamefont {K\"{a}ll},\ and\ \citenamefont
  {B\"{o}rjesson}}]{Xu1999}%
  \BibitemOpen
  \bibfield  {author} {\bibinfo {author} {\bibfnamefont {H.}~\bibnamefont
  {Xu}}, \bibinfo {author} {\bibfnamefont {E.}~\bibnamefont {Bjerneld}},
  \bibinfo {author} {\bibfnamefont {M.}~\bibnamefont {K\"{a}ll}}, \ and\
  \bibinfo {author} {\bibfnamefont {L.}~\bibnamefont {B\"{o}rjesson}},\
  }\href@noop {} {\bibfield  {journal} {\bibinfo  {journal} {Phys. Rev. Lett.}\
  }\textbf {\bibinfo {volume} {83}},\ \bibinfo {pages} {4357} (\bibinfo {year}
  {1999})}\BibitemShut {NoStop}%
\bibitem [{\citenamefont {Baumberg}\ \emph {et~al.}(2019)\citenamefont
  {Baumberg}, \citenamefont {Aizpurua}, \citenamefont {Mikkelsen},\ and\
  \citenamefont {Smith}}]{Buamberg_19}%
  \BibitemOpen
  \bibfield  {author} {\bibinfo {author} {\bibfnamefont {J.~J.}\ \bibnamefont
  {Baumberg}}, \bibinfo {author} {\bibfnamefont {J.}~\bibnamefont {Aizpurua}},
  \bibinfo {author} {\bibfnamefont {M.~H.}\ \bibnamefont {Mikkelsen}}, \ and\
  \bibinfo {author} {\bibfnamefont {D.~R.}\ \bibnamefont {Smith}},\ }\href@noop
  {} {\bibfield  {journal} {\bibinfo  {journal} {Nature Mat.}\ }\textbf
  {\bibinfo {volume} {18}},\ \bibinfo {pages} {668} (\bibinfo {year}
  {2019})}\BibitemShut {NoStop}%
\bibitem [{\citenamefont {Solomon}\ \emph {et~al.}(2010)\citenamefont
  {Solomon}, \citenamefont {Herrmann}, \citenamefont {Hansen}, \citenamefont
  {Mujica},\ and\ \citenamefont {Ratner}}]{Solomon_10}%
  \BibitemOpen
  \bibfield  {author} {\bibinfo {author} {\bibfnamefont {G.~C.}\ \bibnamefont
  {Solomon}}, \bibinfo {author} {\bibfnamefont {C.}~\bibnamefont {Herrmann}},
  \bibinfo {author} {\bibfnamefont {T.}~\bibnamefont {Hansen}}, \bibinfo
  {author} {\bibfnamefont {V.}~\bibnamefont {Mujica}}, \ and\ \bibinfo {author}
  {\bibfnamefont {M.~A.}\ \bibnamefont {Ratner}},\ }\href@noop {} {\bibfield
  {journal} {\bibinfo  {journal} {Nature Chem}\ }\textbf {\bibinfo {volume}
  {2}},\ \bibinfo {pages} {223} (\bibinfo {year} {2010})}\BibitemShut {NoStop}%
\bibitem [{\citenamefont {Ward}\ \emph {et~al.}(2008)\citenamefont {Ward},
  \citenamefont {Halas}, \citenamefont {Ciszek}, \citenamefont {Tour},
  \citenamefont {Wu}, \citenamefont {Nordlander},\ and\ \citenamefont
  {Natelson}}]{Ward2008}%
  \BibitemOpen
  \bibfield  {author} {\bibinfo {author} {\bibfnamefont {D.~R.}\ \bibnamefont
  {Ward}}, \bibinfo {author} {\bibfnamefont {N.~J.}\ \bibnamefont {Halas}},
  \bibinfo {author} {\bibfnamefont {J.~W.}\ \bibnamefont {Ciszek}}, \bibinfo
  {author} {\bibfnamefont {J.~M.}\ \bibnamefont {Tour}}, \bibinfo {author}
  {\bibfnamefont {Y.}~\bibnamefont {Wu}}, \bibinfo {author} {\bibfnamefont
  {P.}~\bibnamefont {Nordlander}}, \ and\ \bibinfo {author} {\bibfnamefont
  {D.}~\bibnamefont {Natelson}},\ }\href@noop {} {\bibfield  {journal}
  {\bibinfo  {journal} {Nano Letters}\ }\textbf {\bibinfo {volume} {8}},\
  \bibinfo {pages} {919} (\bibinfo {year} {2008})}\BibitemShut {NoStop}%
\bibitem [{\citenamefont {Baumberg}(2022)}]{Baumberg_22}%
  \BibitemOpen
  \bibfield  {author} {\bibinfo {author} {\bibfnamefont {J.~J.}\ \bibnamefont
  {Baumberg}},\ }\href@noop {} {\bibfield  {journal} {\bibinfo  {journal} {Nano
  Lett.}\ }\textbf {\bibinfo {volume} {22}},\ \bibinfo {pages} {5859} (\bibinfo
  {year} {2022})}\BibitemShut {NoStop}%
\bibitem [{\citenamefont {Chen}\ \emph {et~al.}(2021)\citenamefont {Chen},
  \citenamefont {Roelli}, \citenamefont {Hu}, \citenamefont {Verlekar},
  \citenamefont {Amirtharaj}, \citenamefont {Barreda}, \citenamefont
  {Kippenberg}, \citenamefont {Kovylina}, \citenamefont {Verhagen},
  \citenamefont {Martinez},\ and\ \citenamefont {Galland}}]{Galland_21}%
  \BibitemOpen
  \bibfield  {author} {\bibinfo {author} {\bibfnamefont {W.}~\bibnamefont
  {Chen}}, \bibinfo {author} {\bibfnamefont {P.}~\bibnamefont {Roelli}},
  \bibinfo {author} {\bibfnamefont {H.}~\bibnamefont {Hu}}, \bibinfo {author}
  {\bibfnamefont {S.}~\bibnamefont {Verlekar}}, \bibinfo {author}
  {\bibfnamefont {S.~P.}\ \bibnamefont {Amirtharaj}}, \bibinfo {author}
  {\bibfnamefont {I.}~\bibnamefont {Barreda}, \bibfnamefont {Angela}}, \bibinfo
  {author} {\bibfnamefont {T.~J.}\ \bibnamefont {Kippenberg}}, \bibinfo
  {author} {\bibfnamefont {M.}~\bibnamefont {Kovylina}}, \bibinfo {author}
  {\bibfnamefont {E.}~\bibnamefont {Verhagen}}, \bibinfo {author}
  {\bibfnamefont {A.}~\bibnamefont {Martinez}}, \ and\ \bibinfo {author}
  {\bibfnamefont {C.}~\bibnamefont {Galland}},\ }\href {\doibase
  10.1126/science.abk3106} {\bibfield  {journal} {\bibinfo  {journal}
  {Science}\ }\textbf {\bibinfo {volume} {374}},\ \bibinfo {pages} {1264}
  (\bibinfo {year} {2021})}\BibitemShut {NoStop}%
\bibitem [{\citenamefont {Esteban}\ \emph {et~al.}(2012)\citenamefont
  {Esteban}, \citenamefont {Borisov}, \citenamefont {Nordlander},\ and\
  \citenamefont {Aizpurua}}]{aizpuruaNC12}%
  \BibitemOpen
  \bibfield  {author} {\bibinfo {author} {\bibfnamefont {R.}~\bibnamefont
  {Esteban}}, \bibinfo {author} {\bibfnamefont {A.~G.}\ \bibnamefont
  {Borisov}}, \bibinfo {author} {\bibfnamefont {P.}~\bibnamefont {Nordlander}},
  \ and\ \bibinfo {author} {\bibfnamefont {J.}~\bibnamefont {Aizpurua}},\
  }\href@noop {} {\bibfield  {journal} {\bibinfo  {journal} {Nat. Commun.}\
  }\textbf {\bibinfo {volume} {3}},\ \bibinfo {pages} {825} (\bibinfo {year}
  {2012})}\BibitemShut {NoStop}%
\bibitem [{\citenamefont {Tame}\ \emph {et~al.}(2013)\citenamefont {Tame},
  \citenamefont {McEnery}, \citenamefont {\"{O}zdemir}, \citenamefont {Lee},
  \citenamefont {Maier},\ and\ \citenamefont {Kim}}]{Tame_13}%
  \BibitemOpen
  \bibfield  {author} {\bibinfo {author} {\bibfnamefont {M.~S.}\ \bibnamefont
  {Tame}}, \bibinfo {author} {\bibfnamefont {K.~R.}\ \bibnamefont {McEnery}},
  \bibinfo {author} {\bibfnamefont {S.~K.}\ \bibnamefont {\"{O}zdemir}},
  \bibinfo {author} {\bibfnamefont {J.}~\bibnamefont {Lee}}, \bibinfo {author}
  {\bibfnamefont {S.~A.}\ \bibnamefont {Maier}}, \ and\ \bibinfo {author}
  {\bibfnamefont {M.~S.}\ \bibnamefont {Kim}},\ }\href@noop {} {\bibfield
  {journal} {\bibinfo  {journal} {Nature Phys.}\ }\textbf {\bibinfo {volume}
  {9}},\ \bibinfo {pages} {329} (\bibinfo {year} {2013})}\BibitemShut {NoStop}%
\bibitem [{\citenamefont {Pang}\ and\ \citenamefont {Gordon}(2012)}]{Pang2012}%
  \BibitemOpen
  \bibfield  {author} {\bibinfo {author} {\bibfnamefont {Y.}~\bibnamefont
  {Pang}}\ and\ \bibinfo {author} {\bibfnamefont {R.}~\bibnamefont {Gordon}},\
  }\href@noop {} {\bibfield  {journal} {\bibinfo  {journal} {Nano Lett.}\
  }\textbf {\bibinfo {volume} {12}},\ \bibinfo {pages} {402} (\bibinfo {year}
  {2012})}\BibitemShut {NoStop}%
\bibitem [{\citenamefont {Berthelot}\ \emph {et~al.}(2014)\citenamefont
  {Berthelot}, \citenamefont {A\'{c}imović}, \citenamefont {Juan},
  \citenamefont {Kreuzer}, \citenamefont {Renger},\ and\ \citenamefont
  {Quidant}}]{Berthelot_14}%
  \BibitemOpen
  \bibfield  {author} {\bibinfo {author} {\bibfnamefont {J.}~\bibnamefont
  {Berthelot}}, \bibinfo {author} {\bibfnamefont {S.~S.}\ \bibnamefont
  {A\'{c}imović}}, \bibinfo {author} {\bibfnamefont {M.~L.}\ \bibnamefont
  {Juan}}, \bibinfo {author} {\bibfnamefont {M.~P.}\ \bibnamefont {Kreuzer}},
  \bibinfo {author} {\bibfnamefont {J.}~\bibnamefont {Renger}}, \ and\ \bibinfo
  {author} {\bibfnamefont {R.}~\bibnamefont {Quidant}},\ }\href@noop {}
  {\bibfield  {journal} {\bibinfo  {journal} {Nature Nanotech.}\ }\textbf
  {\bibinfo {volume} {9}},\ \bibinfo {pages} {295} (\bibinfo {year}
  {2014})}\BibitemShut {NoStop}%
\bibitem [{\citenamefont {Li}\ \emph {et~al.}(2018)\citenamefont {Li},
  \citenamefont {Wang}, \citenamefont {Midya}, \citenamefont {Xia},\ and\
  \citenamefont {Yang}}]{Li_2018}%
  \BibitemOpen
  \bibfield  {author} {\bibinfo {author} {\bibfnamefont {Y.}~\bibnamefont
  {Li}}, \bibinfo {author} {\bibfnamefont {Z.}~\bibnamefont {Wang}}, \bibinfo
  {author} {\bibfnamefont {R.}~\bibnamefont {Midya}}, \bibinfo {author}
  {\bibfnamefont {Q.}~\bibnamefont {Xia}}, \ and\ \bibinfo {author}
  {\bibfnamefont {J.~J.}\ \bibnamefont {Yang}},\ }\href {\doibase
  10.1088/1361-6463/aade3f} {\bibfield  {journal} {\bibinfo  {journal} {Journal
  of Physics D: Applied Physics}\ }\textbf {\bibinfo {volume} {51}},\ \bibinfo
  {pages} {503002} (\bibinfo {year} {2018})}\BibitemShut {NoStop}%
\bibitem [{\citenamefont {Ye}\ \emph {et~al.}(2022)\citenamefont {Ye},
  \citenamefont {Gao}, \citenamefont {Fu}, \citenamefont {Ren}, \citenamefont
  {Yang}, \citenamefont {Wen}, \citenamefont {Wan}, \citenamefont {Ren},
  \citenamefont {Gu}, \citenamefont {Liu}, \citenamefont {Lian},\ and\
  \citenamefont {Wang}}]{Ye_22}%
  \BibitemOpen
  \bibfield  {author} {\bibinfo {author} {\bibfnamefont {L.}~\bibnamefont
  {Ye}}, \bibinfo {author} {\bibfnamefont {Z.}~\bibnamefont {Gao}}, \bibinfo
  {author} {\bibfnamefont {J.}~\bibnamefont {Fu}}, \bibinfo {author}
  {\bibfnamefont {W.}~\bibnamefont {Ren}}, \bibinfo {author} {\bibfnamefont
  {C.}~\bibnamefont {Yang}}, \bibinfo {author} {\bibfnamefont {J.}~\bibnamefont
  {Wen}}, \bibinfo {author} {\bibfnamefont {X.}~\bibnamefont {Wan}}, \bibinfo
  {author} {\bibfnamefont {Q.}~\bibnamefont {Ren}}, \bibinfo {author}
  {\bibfnamefont {S.}~\bibnamefont {Gu}}, \bibinfo {author} {\bibfnamefont
  {X.}~\bibnamefont {Liu}}, \bibinfo {author} {\bibfnamefont {X.}~\bibnamefont
  {Lian}}, \ and\ \bibinfo {author} {\bibfnamefont {L.}~\bibnamefont {Wang}},\
  }\href {\doibase 10.3389/fphy.2022.839243} {\bibfield  {journal} {\bibinfo
  {journal} {Frontiers in Physics}\ }\textbf {\bibinfo {volume} {10}},\
  \bibinfo {pages} {839243} (\bibinfo {year} {2022})}\BibitemShut {NoStop}%
\bibitem [{\citenamefont {Ielmini}\ and\ \citenamefont
  {Ambrogio}(2019)}]{Ambrogio_19}%
  \BibitemOpen
  \bibfield  {author} {\bibinfo {author} {\bibfnamefont {D.}~\bibnamefont
  {Ielmini}}\ and\ \bibinfo {author} {\bibfnamefont {S.}~\bibnamefont
  {Ambrogio}},\ }in\ \href {\doibase
  https://doi.org/10.1016/B978-0-08-102584-0.00017-6} {\emph {\bibinfo
  {booktitle} {Advances in Non-Volatile Memory and Storage Technology (Second
  Edition)}}},\ \bibinfo {series and number} {Woodhead Publishing Series in
  Electronic and Optical Materials},\ \bibinfo {editor} {edited by\ \bibinfo
  {editor} {\bibfnamefont {B.}~\bibnamefont {Magyari-Köpe}}\ and\ \bibinfo
  {editor} {\bibfnamefont {Y.}~\bibnamefont {Nishi}}}\ (\bibinfo  {publisher}
  {Woodhead Publishing},\ \bibinfo {year} {2019})\ \bibinfo {edition} {second
  edition}\ ed.,\ pp.\ \bibinfo {pages} {603--631}\BibitemShut {NoStop}%
\bibitem [{\citenamefont {Tsuruoka}\ \emph {et~al.}(2010)\citenamefont
  {Tsuruoka}, \citenamefont {Terabe}, \citenamefont {Hasegawa},\ and\
  \citenamefont {Aono}}]{Tsuruoka_2010}%
  \BibitemOpen
  \bibfield  {author} {\bibinfo {author} {\bibfnamefont {T.}~\bibnamefont
  {Tsuruoka}}, \bibinfo {author} {\bibfnamefont {K.}~\bibnamefont {Terabe}},
  \bibinfo {author} {\bibfnamefont {T.}~\bibnamefont {Hasegawa}}, \ and\
  \bibinfo {author} {\bibfnamefont {M.}~\bibnamefont {Aono}},\ }\href@noop {}
  {\bibfield  {journal} {\bibinfo  {journal} {Nanotechnology}\ }\textbf
  {\bibinfo {volume} {21}},\ \bibinfo {pages} {425205} (\bibinfo {year}
  {2010})}\BibitemShut {NoStop}%
\bibitem [{\citenamefont {Funck}\ and\ \citenamefont
  {Menzel}(2021)}]{Menzel21}%
  \BibitemOpen
  \bibfield  {author} {\bibinfo {author} {\bibfnamefont {C.}~\bibnamefont
  {Funck}}\ and\ \bibinfo {author} {\bibfnamefont {S.}~\bibnamefont {Menzel}},\
  }\href@noop {} {\bibfield  {journal} {\bibinfo  {journal} {ACS Appl. Elec.
  Mat.}\ }\textbf {\bibinfo {volume} {3}},\ \bibinfo {pages} {3674} (\bibinfo
  {year} {2021})}\BibitemShut {NoStop}%
\bibitem [{\citenamefont {Yao}\ \emph {et~al.}(2012)\citenamefont {Yao},
  \citenamefont {Zhong}, \citenamefont {Natelson},\ and\ \citenamefont
  {Tour}}]{Tour12}%
  \BibitemOpen
  \bibfield  {author} {\bibinfo {author} {\bibfnamefont {J.}~\bibnamefont
  {Yao}}, \bibinfo {author} {\bibfnamefont {L.}~\bibnamefont {Zhong}}, \bibinfo
  {author} {\bibfnamefont {D.}~\bibnamefont {Natelson}}, \ and\ \bibinfo
  {author} {\bibfnamefont {J.~M.}\ \bibnamefont {Tour}},\ }\href@noop {}
  {\bibfield  {journal} {\bibinfo  {journal} {Sci. Rep.}\ }\textbf {\bibinfo
  {volume} {2}},\ \bibinfo {pages} {242} (\bibinfo {year} {2012})}\BibitemShut
  {NoStop}%
\bibitem [{\citenamefont {Yang}\ \emph {et~al.}(2014)\citenamefont {Yang},
  \citenamefont {Gao}, \citenamefont {Li}, \citenamefont {Pan}, \citenamefont
  {Tappertzhofen}, \citenamefont {Choi}, \citenamefont {Waser}, \citenamefont
  {Valov},\ and\ \citenamefont {Lu}}]{Yang2014}%
  \BibitemOpen
  \bibfield  {author} {\bibinfo {author} {\bibfnamefont {Y.}~\bibnamefont
  {Yang}}, \bibinfo {author} {\bibfnamefont {P.}~\bibnamefont {Gao}}, \bibinfo
  {author} {\bibfnamefont {L.}~\bibnamefont {Li}}, \bibinfo {author}
  {\bibfnamefont {X.}~\bibnamefont {Pan}}, \bibinfo {author} {\bibfnamefont
  {S.}~\bibnamefont {Tappertzhofen}}, \bibinfo {author} {\bibfnamefont
  {S.}~\bibnamefont {Choi}}, \bibinfo {author} {\bibfnamefont {R.}~\bibnamefont
  {Waser}}, \bibinfo {author} {\bibfnamefont {I.}~\bibnamefont {Valov}}, \ and\
  \bibinfo {author} {\bibfnamefont {W.~D.}\ \bibnamefont {Lu}},\ }\href@noop {}
  {\bibfield  {journal} {\bibinfo  {journal} {Nature Communications}\ }\textbf
  {\bibinfo {volume} {5}},\ \bibinfo {pages} {4232} (\bibinfo {year}
  {2014})}\BibitemShut {NoStop}%
\bibitem [{\citenamefont {Chang}\ \emph {et~al.}(2014)\citenamefont {Chang},
  \citenamefont {Tan}, \citenamefont {Lu}, \citenamefont {Pan}, \citenamefont
  {Yang},\ and\ \citenamefont {Chen}}]{Chen14}%
  \BibitemOpen
  \bibfield  {author} {\bibinfo {author} {\bibfnamefont {C.-W.}\ \bibnamefont
  {Chang}}, \bibinfo {author} {\bibfnamefont {W.-C.}\ \bibnamefont {Tan}},
  \bibinfo {author} {\bibfnamefont {M.-L.}\ \bibnamefont {Lu}}, \bibinfo
  {author} {\bibfnamefont {T.-C.}\ \bibnamefont {Pan}}, \bibinfo {author}
  {\bibfnamefont {Y.-J.}\ \bibnamefont {Yang}}, \ and\ \bibinfo {author}
  {\bibfnamefont {Y.-F.}\ \bibnamefont {Chen}},\ }\href@noop {} {\bibfield
  {journal} {\bibinfo  {journal} {Sci. Rep.}\ }\textbf {\bibinfo {volume}
  {4}},\ \bibinfo {pages} {5121} (\bibinfo {year} {2014})}\BibitemShut
  {NoStop}%
\bibitem [{\citenamefont {Schoen}\ \emph {et~al.}(2016)\citenamefont {Schoen},
  \citenamefont {Holsteen},\ and\ \citenamefont {Brongersma}}]{Brongersma16}%
  \BibitemOpen
  \bibfield  {author} {\bibinfo {author} {\bibfnamefont {D.~T.}\ \bibnamefont
  {Schoen}}, \bibinfo {author} {\bibfnamefont {A.~L.}\ \bibnamefont
  {Holsteen}}, \ and\ \bibinfo {author} {\bibfnamefont {M.~L.}\ \bibnamefont
  {Brongersma}},\ }\href@noop {} {\bibfield  {journal} {\bibinfo  {journal}
  {Nat. Commun.}\ }\textbf {\bibinfo {volume} {7}},\ \bibinfo {pages} {12162}
  (\bibinfo {year} {2016})}\BibitemShut {NoStop}%
\bibitem [{\citenamefont {Emboras}\ \emph {et~al.}(2016)\citenamefont
  {Emboras}, \citenamefont {Niegemann}, \citenamefont {Ma}, \citenamefont
  {Haffner}, \citenamefont {Pedersen}, \citenamefont {Luisier}, \citenamefont
  {Hafner}, \citenamefont {Schimmel},\ and\ \citenamefont
  {Leuthold}}]{Emboras16}%
  \BibitemOpen
  \bibfield  {author} {\bibinfo {author} {\bibfnamefont {A.}~\bibnamefont
  {Emboras}}, \bibinfo {author} {\bibfnamefont {J.}~\bibnamefont {Niegemann}},
  \bibinfo {author} {\bibfnamefont {P.}~\bibnamefont {Ma}}, \bibinfo {author}
  {\bibfnamefont {C.}~\bibnamefont {Haffner}}, \bibinfo {author} {\bibfnamefont
  {A.}~\bibnamefont {Pedersen}}, \bibinfo {author} {\bibfnamefont
  {M.}~\bibnamefont {Luisier}}, \bibinfo {author} {\bibfnamefont
  {C.}~\bibnamefont {Hafner}}, \bibinfo {author} {\bibfnamefont
  {T.}~\bibnamefont {Schimmel}}, \ and\ \bibinfo {author} {\bibfnamefont
  {J.}~\bibnamefont {Leuthold}},\ }\href@noop {} {\bibfield  {journal}
  {\bibinfo  {journal} {Nano Lett.}\ }\textbf {\bibinfo {volume} {16}},\
  \bibinfo {pages} {709} (\bibinfo {year} {2016})}\BibitemShut {NoStop}%
\bibitem [{\citenamefont {Koch}\ \emph {et~al.}(2017)\citenamefont {Koch},
  \citenamefont {Hoessbacher}, \citenamefont {Emboras},\ and\ \citenamefont
  {Leuthold}}]{Leuthold17}%
  \BibitemOpen
  \bibfield  {author} {\bibinfo {author} {\bibfnamefont {U.}~\bibnamefont
  {Koch}}, \bibinfo {author} {\bibfnamefont {C.}~\bibnamefont {Hoessbacher}},
  \bibinfo {author} {\bibfnamefont {A.}~\bibnamefont {Emboras}}, \ and\
  \bibinfo {author} {\bibfnamefont {J.}~\bibnamefont {Leuthold}},\ }\href@noop
  {} {\bibfield  {journal} {\bibinfo  {journal} {Journal of Electroceramics}\
  }\textbf {\bibinfo {volume} {39}},\ \bibinfo {pages} {239} (\bibinfo {year}
  {2017})}\BibitemShut {NoStop}%
\bibitem [{\citenamefont {Tang}\ \emph {et~al.}(2021)\citenamefont {Tang},
  \citenamefont {Hu}, \citenamefont {He}, \citenamefont {Xu}, \citenamefont
  {Zhang}, \citenamefont {Guan}, \citenamefont {Zhang},\ and\ \citenamefont
  {Xu}}]{Xu21}%
  \BibitemOpen
  \bibfield  {author} {\bibinfo {author} {\bibfnamefont {J.}~\bibnamefont
  {Tang}}, \bibinfo {author} {\bibfnamefont {H.}~\bibnamefont {Hu}}, \bibinfo
  {author} {\bibfnamefont {X.}~\bibnamefont {He}}, \bibinfo {author}
  {\bibfnamefont {Y.}~\bibnamefont {Xu}}, \bibinfo {author} {\bibfnamefont
  {Y.}~\bibnamefont {Zhang}}, \bibinfo {author} {\bibfnamefont
  {Z.}~\bibnamefont {Guan}}, \bibinfo {author} {\bibfnamefont {S.}~\bibnamefont
  {Zhang}}, \ and\ \bibinfo {author} {\bibfnamefont {H.}~\bibnamefont {Xu}},\
  }\href@noop {} {\bibfield  {journal} {\bibinfo  {journal} {Adv. Opt. Mat.}\
  }\textbf {\bibinfo {volume} {9}},\ \bibinfo {pages} {2100191} (\bibinfo
  {year} {2021})}\BibitemShut {NoStop}%
\bibitem [{\citenamefont {Kern}\ \emph {et~al.}(2015)\citenamefont {Kern},
  \citenamefont {Kullock}, \citenamefont {Prangsma}, \citenamefont {Emmerling},
  \citenamefont {Kamp},\ and\ \citenamefont {Hecht}}]{Hecht15}%
  \BibitemOpen
  \bibfield  {author} {\bibinfo {author} {\bibfnamefont {J.}~\bibnamefont
  {Kern}}, \bibinfo {author} {\bibfnamefont {R.}~\bibnamefont {Kullock}},
  \bibinfo {author} {\bibfnamefont {J.~C.}\ \bibnamefont {Prangsma}}, \bibinfo
  {author} {\bibfnamefont {M.}~\bibnamefont {Emmerling}}, \bibinfo {author}
  {\bibfnamefont {M.}~\bibnamefont {Kamp}}, \ and\ \bibinfo {author}
  {\bibfnamefont {B.}~\bibnamefont {Hecht}},\ }\href@noop {} {\bibfield
  {journal} {\bibinfo  {journal} {Nat. Photonics}\ }\textbf {\bibinfo {volume}
  {9}},\ \bibinfo {pages} {582} (\bibinfo {year} {2015})}\BibitemShut {NoStop}%
\bibitem [{\citenamefont {Parzefall}\ and\ \citenamefont
  {Novotny}(2019)}]{Parzefall_2019}%
  \BibitemOpen
  \bibfield  {author} {\bibinfo {author} {\bibfnamefont {M.}~\bibnamefont
  {Parzefall}}\ and\ \bibinfo {author} {\bibfnamefont {L.}~\bibnamefont
  {Novotny}},\ }\href {\doibase 10.1088/1361-6633/ab4239} {\bibfield  {journal}
  {\bibinfo  {journal} {Reports on Progress in Physics}\ }\textbf {\bibinfo
  {volume} {82}},\ \bibinfo {pages} {112401} (\bibinfo {year}
  {2019})}\BibitemShut {NoStop}%
\bibitem [{\citenamefont {Cheng}\ \emph {et~al.}(2022)\citenamefont {Cheng},
  \citenamefont {Zellweger}, \citenamefont {Malchow}, \citenamefont {Zhang},
  \citenamefont {Lewerenz}, \citenamefont {Passerini}, \citenamefont
  {Aeschlimann}, \citenamefont {Koch}, \citenamefont {Luisier}, \citenamefont
  {Emboras}, \citenamefont {Bouhelier},\ and\ \citenamefont
  {Leuthold}}]{Cheng2022}%
  \BibitemOpen
  \bibfield  {author} {\bibinfo {author} {\bibfnamefont {B.}~\bibnamefont
  {Cheng}}, \bibinfo {author} {\bibfnamefont {T.}~\bibnamefont {Zellweger}},
  \bibinfo {author} {\bibfnamefont {K.}~\bibnamefont {Malchow}}, \bibinfo
  {author} {\bibfnamefont {X.}~\bibnamefont {Zhang}}, \bibinfo {author}
  {\bibfnamefont {M.}~\bibnamefont {Lewerenz}}, \bibinfo {author}
  {\bibfnamefont {E.}~\bibnamefont {Passerini}}, \bibinfo {author}
  {\bibfnamefont {J.}~\bibnamefont {Aeschlimann}}, \bibinfo {author}
  {\bibfnamefont {U.}~\bibnamefont {Koch}}, \bibinfo {author} {\bibfnamefont
  {M.}~\bibnamefont {Luisier}}, \bibinfo {author} {\bibfnamefont
  {A.}~\bibnamefont {Emboras}}, \bibinfo {author} {\bibfnamefont
  {A.}~\bibnamefont {Bouhelier}}, \ and\ \bibinfo {author} {\bibfnamefont
  {J.}~\bibnamefont {Leuthold}},\ }\href@noop {} {\bibfield  {journal}
  {\bibinfo  {journal} {Light: Science \& Applications}\ }\textbf {\bibinfo
  {volume} {11}},\ \bibinfo {pages} {78} (\bibinfo {year} {2022})}\BibitemShut
  {NoStop}%
\bibitem [{\citenamefont {Tomchuk}\ and\ \citenamefont
  {Fedorovich}(1966)}]{Tomchuk66}%
  \BibitemOpen
  \bibfield  {author} {\bibinfo {author} {\bibfnamefont {P.}~\bibnamefont
  {Tomchuk}}\ and\ \bibinfo {author} {\bibfnamefont {R.}~\bibnamefont
  {Fedorovich}},\ }\href@noop {} {\bibfield  {journal} {\bibinfo  {journal}
  {Sov. Phys. Sol. Stat.}\ ,\ \bibinfo {pages} {276}} (\bibinfo {year}
  {1966})}\BibitemShut {NoStop}%
\bibitem [{\citenamefont {Buret}\ \emph {et~al.}(2015)\citenamefont {Buret},
  \citenamefont {Uskov}, \citenamefont {Dellinger}, \citenamefont {Cazier},
  \citenamefont {Mennemanteuil}, \citenamefont {Berthelot}, \citenamefont
  {Smetanin}, \citenamefont {Protsenko}, \citenamefont {{Colas-des-Francs}},\
  and\ \citenamefont {Bouhelier}}]{Buret2015}%
  \BibitemOpen
  \bibfield  {author} {\bibinfo {author} {\bibfnamefont {M.}~\bibnamefont
  {Buret}}, \bibinfo {author} {\bibfnamefont {A.~V.}\ \bibnamefont {Uskov}},
  \bibinfo {author} {\bibfnamefont {J.}~\bibnamefont {Dellinger}}, \bibinfo
  {author} {\bibfnamefont {N.}~\bibnamefont {Cazier}}, \bibinfo {author}
  {\bibfnamefont {M.-M.}\ \bibnamefont {Mennemanteuil}}, \bibinfo {author}
  {\bibfnamefont {J.}~\bibnamefont {Berthelot}}, \bibinfo {author}
  {\bibfnamefont {I.~V.}\ \bibnamefont {Smetanin}}, \bibinfo {author}
  {\bibfnamefont {I.~E.}\ \bibnamefont {Protsenko}}, \bibinfo {author}
  {\bibfnamefont {G.}~\bibnamefont {{Colas-des-Francs}}}, \ and\ \bibinfo
  {author} {\bibfnamefont {A.}~\bibnamefont {Bouhelier}},\ }\href@noop {}
  {\bibfield  {journal} {\bibinfo  {journal} {Nano Lett.}\ }\textbf {\bibinfo
  {volume} {15}},\ \bibinfo {pages} {5811} (\bibinfo {year}
  {2015})}\BibitemShut {NoStop}%
\bibitem [{\citenamefont {Cui}\ \emph {et~al.}(2020)\citenamefont {Cui},
  \citenamefont {Zhu}, \citenamefont {Abbasi}, \citenamefont {Ahmadivand},
  \citenamefont {Gerislioglu}, \citenamefont {Nordlander},\ and\ \citenamefont
  {Natelson}}]{Cui20}%
  \BibitemOpen
  \bibfield  {author} {\bibinfo {author} {\bibfnamefont {L.}~\bibnamefont
  {Cui}}, \bibinfo {author} {\bibfnamefont {Y.}~\bibnamefont {Zhu}}, \bibinfo
  {author} {\bibfnamefont {M.}~\bibnamefont {Abbasi}}, \bibinfo {author}
  {\bibfnamefont {A.}~\bibnamefont {Ahmadivand}}, \bibinfo {author}
  {\bibfnamefont {B.}~\bibnamefont {Gerislioglu}}, \bibinfo {author}
  {\bibfnamefont {P.}~\bibnamefont {Nordlander}}, \ and\ \bibinfo {author}
  {\bibfnamefont {D.}~\bibnamefont {Natelson}},\ }\href@noop {} {\bibfield
  {journal} {\bibinfo  {journal} {Nano Lett.}\ }\textbf {\bibinfo {volume}
  {20}},\ \bibinfo {pages} {6067} (\bibinfo {year} {2020})}\BibitemShut
  {NoStop}%
\bibitem [{\citenamefont {Malinowski}\ \emph {et~al.}(2016)\citenamefont
  {Malinowski}, \citenamefont {Klein}, \citenamefont {Iazykov},\ and\
  \citenamefont {Dumas}}]{Dumas16}%
  \BibitemOpen
  \bibfield  {author} {\bibinfo {author} {\bibfnamefont {T.}~\bibnamefont
  {Malinowski}}, \bibinfo {author} {\bibfnamefont {H.~R.}\ \bibnamefont
  {Klein}}, \bibinfo {author} {\bibfnamefont {M.}~\bibnamefont {Iazykov}}, \
  and\ \bibinfo {author} {\bibfnamefont {P.}~\bibnamefont {Dumas}},\
  }\href@noop {} {\bibfield  {journal} {\bibinfo  {journal} {Europhys. Lett.}\
  }\textbf {\bibinfo {volume} {114}},\ \bibinfo {pages} {57002} (\bibinfo
  {year} {2016})}\BibitemShut {NoStop}%
\bibitem [{\citenamefont {Downes}\ \emph {et~al.}(2002)\citenamefont {Downes},
  \citenamefont {Dumas},\ and\ \citenamefont {Welland}}]{welland02}%
  \BibitemOpen
  \bibfield  {author} {\bibinfo {author} {\bibfnamefont {A.}~\bibnamefont
  {Downes}}, \bibinfo {author} {\bibfnamefont {P.}~\bibnamefont {Dumas}}, \
  and\ \bibinfo {author} {\bibfnamefont {M.~E.}\ \bibnamefont {Welland}},\
  }\href@noop {} {\bibfield  {journal} {\bibinfo  {journal} {Appl. Phys.
  Lett.}\ }\textbf {\bibinfo {volume} {81}},\ \bibinfo {pages} {1252} (\bibinfo
  {year} {2002})}\BibitemShut {NoStop}%
\bibitem [{\citenamefont {Peters}\ \emph {et~al.}(2017)\citenamefont {Peters},
  \citenamefont {Xu}, \citenamefont {Kaasbjerg}, \citenamefont {Rastelli},
  \citenamefont {Belzig},\ and\ \citenamefont {Berndt}}]{Berndt17}%
  \BibitemOpen
  \bibfield  {author} {\bibinfo {author} {\bibfnamefont {P.-J.}\ \bibnamefont
  {Peters}}, \bibinfo {author} {\bibfnamefont {F.}~\bibnamefont {Xu}}, \bibinfo
  {author} {\bibfnamefont {K.}~\bibnamefont {Kaasbjerg}}, \bibinfo {author}
  {\bibfnamefont {G.}~\bibnamefont {Rastelli}}, \bibinfo {author}
  {\bibfnamefont {W.}~\bibnamefont {Belzig}}, \ and\ \bibinfo {author}
  {\bibfnamefont {R.}~\bibnamefont {Berndt}},\ }\href {\doibase
  10.1103/PhysRevLett.119.066803} {\bibfield  {journal} {\bibinfo  {journal}
  {Phys. Rev. Lett.}\ }\textbf {\bibinfo {volume} {119}},\ \bibinfo {pages}
  {066803} (\bibinfo {year} {2017})}\BibitemShut {NoStop}%
\bibitem [{\citenamefont {Wang}\ \emph {et~al.}(2017)\citenamefont {Wang},
  \citenamefont {Joshi}, \citenamefont {Savel'ev}, \citenamefont {Jiang},
  \citenamefont {Midya}, \citenamefont {Lin}, \citenamefont {Hu}, \citenamefont
  {Ge}, \citenamefont {Strachan}, \citenamefont {Li}, \citenamefont {Wu},
  \citenamefont {Barnell}, \citenamefont {Li}, \citenamefont {Xin},
  \citenamefont {Williams}, \citenamefont {Xia},\ and\ \citenamefont
  {Yang}}]{Wang17}%
  \BibitemOpen
  \bibfield  {author} {\bibinfo {author} {\bibfnamefont {Z.}~\bibnamefont
  {Wang}}, \bibinfo {author} {\bibfnamefont {S.}~\bibnamefont {Joshi}},
  \bibinfo {author} {\bibfnamefont {S.~E.}\ \bibnamefont {Savel'ev}}, \bibinfo
  {author} {\bibfnamefont {H.}~\bibnamefont {Jiang}}, \bibinfo {author}
  {\bibfnamefont {R.}~\bibnamefont {Midya}}, \bibinfo {author} {\bibfnamefont
  {P.}~\bibnamefont {Lin}}, \bibinfo {author} {\bibfnamefont {M.}~\bibnamefont
  {Hu}}, \bibinfo {author} {\bibfnamefont {N.}~\bibnamefont {Ge}}, \bibinfo
  {author} {\bibfnamefont {J.~P.}\ \bibnamefont {Strachan}}, \bibinfo {author}
  {\bibfnamefont {Z.}~\bibnamefont {Li}}, \bibinfo {author} {\bibfnamefont
  {Q.}~\bibnamefont {Wu}}, \bibinfo {author} {\bibfnamefont {M.}~\bibnamefont
  {Barnell}}, \bibinfo {author} {\bibfnamefont {G.-L.}\ \bibnamefont {Li}},
  \bibinfo {author} {\bibfnamefont {H.~L.}\ \bibnamefont {Xin}}, \bibinfo
  {author} {\bibfnamefont {R.~S.}\ \bibnamefont {Williams}}, \bibinfo {author}
  {\bibfnamefont {Q.}~\bibnamefont {Xia}}, \ and\ \bibinfo {author}
  {\bibfnamefont {J.~J.}\ \bibnamefont {Yang}},\ }\href {\doibase
  10.1038/nmat4756} {\bibfield  {journal} {\bibinfo  {journal} {Nature
  Materials}\ }\textbf {\bibinfo {volume} {16}},\ \bibinfo {pages} {101}
  (\bibinfo {year} {2017})}\BibitemShut {NoStop}%
\bibitem [{\citenamefont {Patel}\ \emph {et~al.}(2019)\citenamefont {Patel},
  \citenamefont {Cottom}, \citenamefont {Bosman}, \citenamefont {Kenyon},\ and\
  \citenamefont {Shluger}}]{Patel19}%
  \BibitemOpen
  \bibfield  {author} {\bibinfo {author} {\bibfnamefont {K.}~\bibnamefont
  {Patel}}, \bibinfo {author} {\bibfnamefont {J.}~\bibnamefont {Cottom}},
  \bibinfo {author} {\bibfnamefont {M.}~\bibnamefont {Bosman}}, \bibinfo
  {author} {\bibfnamefont {A.}~\bibnamefont {Kenyon}}, \ and\ \bibinfo {author}
  {\bibfnamefont {A.}~\bibnamefont {Shluger}},\ }\href {\doibase
  https://doi.org/10.1016/j.microrel.2019.05.005} {\bibfield  {journal}
  {\bibinfo  {journal} {Microelectronics Reliability}\ }\textbf {\bibinfo
  {volume} {98}},\ \bibinfo {pages} {144} (\bibinfo {year} {2019})}\BibitemShut
  {NoStop}%
\bibitem [{\citenamefont {Michaelis}\ \emph {et~al.}(2000)\citenamefont
  {Michaelis}, \citenamefont {Hettich}, \citenamefont {Mlynek},\ and\
  \citenamefont {Sandoghdar}}]{michaelis2000}%
  \BibitemOpen
  \bibfield  {author} {\bibinfo {author} {\bibfnamefont {J.}~\bibnamefont
  {Michaelis}}, \bibinfo {author} {\bibfnamefont {C.}~\bibnamefont {Hettich}},
  \bibinfo {author} {\bibfnamefont {J.}~\bibnamefont {Mlynek}}, \ and\ \bibinfo
  {author} {\bibfnamefont {V.}~\bibnamefont {Sandoghdar}},\ }\href@noop {}
  {\bibfield  {journal} {\bibinfo  {journal} {Nature}\ }\textbf {\bibinfo
  {volume} {405}},\ \bibinfo {pages} {325} (\bibinfo {year}
  {2000})}\BibitemShut {NoStop}%
\bibitem [{\citenamefont {Cheng}\ \emph {et~al.}(2019)\citenamefont {Cheng},
  \citenamefont {Emboras}, \citenamefont {Salamin}, \citenamefont {Ducry},
  \citenamefont {Ma}, \citenamefont {Fedoryshyn}, \citenamefont {Andermatt},
  \citenamefont {Luisier},\ and\ \citenamefont {Leuthold}}]{Cheng_19}%
  \BibitemOpen
  \bibfield  {author} {\bibinfo {author} {\bibfnamefont {B.}~\bibnamefont
  {Cheng}}, \bibinfo {author} {\bibfnamefont {A.}~\bibnamefont {Emboras}},
  \bibinfo {author} {\bibfnamefont {Y.}~\bibnamefont {Salamin}}, \bibinfo
  {author} {\bibfnamefont {F.}~\bibnamefont {Ducry}}, \bibinfo {author}
  {\bibfnamefont {P.}~\bibnamefont {Ma}}, \bibinfo {author} {\bibfnamefont
  {Y.}~\bibnamefont {Fedoryshyn}}, \bibinfo {author} {\bibfnamefont
  {S.}~\bibnamefont {Andermatt}}, \bibinfo {author} {\bibfnamefont
  {M.}~\bibnamefont {Luisier}}, \ and\ \bibinfo {author} {\bibfnamefont
  {J.}~\bibnamefont {Leuthold}},\ }\href@noop {} {\bibfield  {journal}
  {\bibinfo  {journal} {Communications Physics}\ }\textbf {\bibinfo {volume}
  {2}},\ \bibinfo {pages} {28} (\bibinfo {year} {2019})}\BibitemShut {NoStop}%
\bibitem [{\citenamefont {Pechou}\ \emph {et~al.}(1998)\citenamefont {Pechou},
  \citenamefont {Coratger}, \citenamefont {Ajustron},\ and\ \citenamefont
  {Beauvillain}}]{pechou98}%
  \BibitemOpen
  \bibfield  {author} {\bibinfo {author} {\bibfnamefont {R.}~\bibnamefont
  {Pechou}}, \bibinfo {author} {\bibfnamefont {R.}~\bibnamefont {Coratger}},
  \bibinfo {author} {\bibfnamefont {F.}~\bibnamefont {Ajustron}}, \ and\
  \bibinfo {author} {\bibfnamefont {J.}~\bibnamefont {Beauvillain}},\
  }\href@noop {} {\bibfield  {journal} {\bibinfo  {journal} {Appl. Phys.
  Lett.}\ }\textbf {\bibinfo {volume} {72}},\ \bibinfo {pages} {671} (\bibinfo
  {year} {1998})}\BibitemShut {NoStop}%
\bibitem [{\citenamefont {Schull}\ \emph {et~al.}(2009)\citenamefont {Schull},
  \citenamefont {N\'{e}el}, \citenamefont {Johansson},\ and\ \citenamefont
  {Berndt}}]{schull09}%
  \BibitemOpen
  \bibfield  {author} {\bibinfo {author} {\bibfnamefont {G.}~\bibnamefont
  {Schull}}, \bibinfo {author} {\bibfnamefont {N.}~\bibnamefont {N\'{e}el}},
  \bibinfo {author} {\bibfnamefont {P.}~\bibnamefont {Johansson}}, \ and\
  \bibinfo {author} {\bibfnamefont {R.}~\bibnamefont {Berndt}},\ }\href@noop {}
  {\bibfield  {journal} {\bibinfo  {journal} {Phys. Rev. Lett.}\ }\textbf
  {\bibinfo {volume} {102}},\ \bibinfo {pages} {057401} (\bibinfo {year}
  {2009})}\BibitemShut {NoStop}%
\bibitem [{\citenamefont {Schneider}\ \emph {et~al.}(2013)\citenamefont
  {Schneider}, \citenamefont {Johansson},\ and\ \citenamefont
  {Berndt}}]{Schneider13}%
  \BibitemOpen
  \bibfield  {author} {\bibinfo {author} {\bibfnamefont {N.~L.}\ \bibnamefont
  {Schneider}}, \bibinfo {author} {\bibfnamefont {P.}~\bibnamefont
  {Johansson}}, \ and\ \bibinfo {author} {\bibfnamefont {R.}~\bibnamefont
  {Berndt}},\ }\href@noop {} {\bibfield  {journal} {\bibinfo  {journal} {Phys.
  Rev. B}\ }\textbf {\bibinfo {volume} {87}},\ \bibinfo {pages} {045409}
  (\bibinfo {year} {2013})}\BibitemShut {NoStop}%
\bibitem [{\citenamefont {Xu}\ \emph {et~al.}(2014)\citenamefont {Xu},
  \citenamefont {Holmqvist},\ and\ \citenamefont {Belzig}}]{belzig14}%
  \BibitemOpen
  \bibfield  {author} {\bibinfo {author} {\bibfnamefont {F.}~\bibnamefont
  {Xu}}, \bibinfo {author} {\bibfnamefont {C.}~\bibnamefont {Holmqvist}}, \
  and\ \bibinfo {author} {\bibfnamefont {W.}~\bibnamefont {Belzig}},\
  }\href@noop {} {\bibfield  {journal} {\bibinfo  {journal} {Phys. Rev. Lett.}\
  }\textbf {\bibinfo {volume} {113}},\ \bibinfo {pages} {066801} (\bibinfo
  {year} {2014})}\BibitemShut {NoStop}%
\bibitem [{\citenamefont {Kaasbjerg}\ and\ \citenamefont
  {Nitzan}(2015)}]{Nitzan15}%
  \BibitemOpen
  \bibfield  {author} {\bibinfo {author} {\bibfnamefont {K.}~\bibnamefont
  {Kaasbjerg}}\ and\ \bibinfo {author} {\bibfnamefont {A.}~\bibnamefont
  {Nitzan}},\ }\href {\doibase 10.1103/PhysRevLett.114.126803} {\bibfield
  {journal} {\bibinfo  {journal} {Phys. Rev. Lett.}\ }\textbf {\bibinfo
  {volume} {114}},\ \bibinfo {pages} {126803} (\bibinfo {year}
  {2015})}\BibitemShut {NoStop}%
\bibitem [{\citenamefont {Fung}\ and\ \citenamefont
  {Venkataraman}(2020)}]{Venkataraman20}%
  \BibitemOpen
  \bibfield  {author} {\bibinfo {author} {\bibfnamefont {E.-D.}\ \bibnamefont
  {Fung}}\ and\ \bibinfo {author} {\bibfnamefont {L.}~\bibnamefont
  {Venkataraman}},\ }\href@noop {} {\bibfield  {journal} {\bibinfo  {journal}
  {Nano Letters}\ }\textbf {\bibinfo {volume} {20}},\ \bibinfo {pages} {8912}
  (\bibinfo {year} {2020})}\BibitemShut {NoStop}%
\bibitem [{\citenamefont {Zhu}\ \emph {et~al.}(2020)\citenamefont {Zhu},
  \citenamefont {Cui},\ and\ \citenamefont {Natelson}}]{Zhu2020}%
  \BibitemOpen
  \bibfield  {author} {\bibinfo {author} {\bibfnamefont {Y.}~\bibnamefont
  {Zhu}}, \bibinfo {author} {\bibfnamefont {L.}~\bibnamefont {Cui}}, \ and\
  \bibinfo {author} {\bibfnamefont {D.}~\bibnamefont {Natelson}},\ }\href@noop
  {} {\bibfield  {journal} {\bibinfo  {journal} {J. Appl. Phys.}\ }\textbf
  {\bibinfo {volume} {128}},\ \bibinfo {pages} {233105} (\bibinfo {year}
  {2020})}\BibitemShut {NoStop}%
\bibitem [{\citenamefont {Chopra}\ \emph {et~al.}(1963)\citenamefont {Chopra},
  \citenamefont {Bobb},\ and\ \citenamefont {Francombe}}]{Chopra63}%
  \BibitemOpen
  \bibfield  {author} {\bibinfo {author} {\bibfnamefont {K.~L.}\ \bibnamefont
  {Chopra}}, \bibinfo {author} {\bibfnamefont {L.~C.}\ \bibnamefont {Bobb}}, \
  and\ \bibinfo {author} {\bibfnamefont {M.~H.}\ \bibnamefont {Francombe}},\
  }\href {\doibase http://dx.doi.org/10.1063/1.1702662} {\bibfield  {journal}
  {\bibinfo  {journal} {J. Appl. Phys.}\ }\textbf {\bibinfo {volume} {34}},\
  \bibinfo {pages} {1699} (\bibinfo {year} {1963})}\BibitemShut {NoStop}%
\bibitem [{\citenamefont {Sze}\ \emph {et~al.}(1964)\citenamefont {Sze},
  \citenamefont {Moll},\ and\ \citenamefont {Sugano}}]{Sze64}%
  \BibitemOpen
  \bibfield  {author} {\bibinfo {author} {\bibfnamefont {S.}~\bibnamefont
  {Sze}}, \bibinfo {author} {\bibfnamefont {J.}~\bibnamefont {Moll}}, \ and\
  \bibinfo {author} {\bibfnamefont {T.}~\bibnamefont {Sugano}},\ }\href
  {\doibase https://doi.org/10.1016/0038-1101(64)90088-7} {\bibfield  {journal}
  {\bibinfo  {journal} {Solid-State Electronics}\ }\textbf {\bibinfo {volume}
  {7}},\ \bibinfo {pages} {509} (\bibinfo {year} {1964})}\BibitemShut {NoStop}%
\bibitem [{\citenamefont {Fedorovich}\ \emph {et~al.}(2000)\citenamefont
  {Fedorovich}, \citenamefont {Naumovets},\ and\ \citenamefont
  {Tomchuk}}]{Federovich00}%
  \BibitemOpen
  \bibfield  {author} {\bibinfo {author} {\bibfnamefont {R.}~\bibnamefont
  {Fedorovich}}, \bibinfo {author} {\bibfnamefont {A.}~\bibnamefont
  {Naumovets}}, \ and\ \bibinfo {author} {\bibfnamefont {P.}~\bibnamefont
  {Tomchuk}},\ }\href@noop {} {\bibfield  {journal} {\bibinfo  {journal} {Phys.
  Rep.}\ }\textbf {\bibinfo {volume} {328}},\ \bibinfo {pages} {73} (\bibinfo
  {year} {2000})}\BibitemShut {NoStop}%
\bibitem [{\citenamefont {Buret}\ \emph {et~al.}(2020)\citenamefont {Buret},
  \citenamefont {Smetanin}, \citenamefont {Uskov}, \citenamefont {{Colas des
  Francs}},\ and\ \citenamefont {Bouhelier}}]{Buret20}%
  \BibitemOpen
  \bibfield  {author} {\bibinfo {author} {\bibfnamefont {M.}~\bibnamefont
  {Buret}}, \bibinfo {author} {\bibfnamefont {I.~V.}\ \bibnamefont {Smetanin}},
  \bibinfo {author} {\bibfnamefont {A.~V.}\ \bibnamefont {Uskov}}, \bibinfo
  {author} {\bibfnamefont {G.}~\bibnamefont {{Colas des Francs}}}, \ and\
  \bibinfo {author} {\bibfnamefont {A.}~\bibnamefont {Bouhelier}},\ }\href@noop
  {} {\bibfield  {journal} {\bibinfo  {journal} {Nanophot.}\ }\textbf {\bibinfo
  {volume} {9}},\ \bibinfo {pages} {413} (\bibinfo {year} {2020})}\BibitemShut
  {NoStop}%
\bibitem [{\citenamefont {Ghisellini}(2013)}]{Ghisellini}%
  \BibitemOpen
  \bibfield  {author} {\bibinfo {author} {\bibfnamefont {G.}~\bibnamefont
  {Ghisellini}},\ }\href@noop {} {\emph {\bibinfo {title} {Radiative Processes
  in High Energy Astrophysics. Lecture Notes in Physics,}}}\ (\bibinfo
  {publisher} {Springer, Heidelberg},\ \bibinfo {year} {2013})\BibitemShut
  {NoStop}%
\bibitem [{\citenamefont {Zhu}\ \emph {et~al.}(2022)\citenamefont {Zhu},
  \citenamefont {Cui}, \citenamefont {Abbasi},\ and\ \citenamefont
  {Natelson}}]{Natelson22}%
  \BibitemOpen
  \bibfield  {author} {\bibinfo {author} {\bibfnamefont {Y.}~\bibnamefont
  {Zhu}}, \bibinfo {author} {\bibfnamefont {L.}~\bibnamefont {Cui}}, \bibinfo
  {author} {\bibfnamefont {M.}~\bibnamefont {Abbasi}}, \ and\ \bibinfo {author}
  {\bibfnamefont {D.}~\bibnamefont {Natelson}},\ }\href@noop {} {\bibfield
  {journal} {\bibinfo  {journal} {Nano Letters}\ }\textbf {\bibinfo {volume}
  {22}},\ \bibinfo {pages} {8068} (\bibinfo {year} {2022})}\BibitemShut
  {NoStop}%
\end{thebibliography}%

\end{document}